\documentclass{article}

\usepackage{algorithm}
\usepackage[margin=30mm,includehead,includefoot]{geometry}
\usepackage[noend]{algpseudocode}
\usepackage{amsmath,amssymb,amsthm,amscd,color,comment}
\usepackage{autobreak}
\usepackage{color}
\usepackage{xcolor}
\usepackage{url}
\usepackage{blkarray}
\usepackage{jheppub}
\usepackage{empheq}
\usepackage{graphicx}
\usepackage{booktabs, multirow}
\usepackage{subcaption}
\usepackage{bm}
\usepackage{enumerate}

\usepackage{tikz}
\usetikzlibrary{patterns}
\usetikzlibrary{arrows,shapes,positioning}
\usetikzlibrary{trees}
\usetikzlibrary{matrix,arrows} 				% For commutative diagram
\usetikzlibrary{positioning}				% For "above of=" commands
\usetikzlibrary{calc,through}				% For coordinates
\usetikzlibrary{decorations.pathreplacing}  % For curly braces
\usepackage{pgffor}							% For repeating patterns
\usetikzlibrary{decorations.pathmorphing}	% For Feynman Diagrams
\usetikzlibrary{decorations.markings}
\tikzstyle{block} = [draw, rectangle, 
minimum height=3em, minimum width=6em]

\tikzstyle{decision} = [diamond, draw, fill=blue!20, 
text width=4.5em, text badly centered, node distance=3cm, inner sep=0pt]
\tikzstyle{block} = [rectangle, draw, fill=blue!10, 
text width=7em, text centered, rounded corners, minimum height=4em, node distance=3cm]
\tikzstyle{line} = [draw, -latex']
\tikzstyle{cloud} = [draw, rectangle, fill=blue!20, text width=6em, text centered, rounded corners, node distance=3cm,
minimum height=2em]
\tikzstyle{border} = [draw, dashed, rectangle, fill=blue!5, rounded corners, node distance=3cm, minimum height=25em, minimum width=22em]
\tikzstyle{data} = [draw, rectangle, fill=blue!10, rounded corners, minimum height=2em, minimum width=8em]
\tikzstyle{box} = [draw, rectangle, fill=white, rounded corners, minimum width=10em]
\tikzstyle{input}=[trapezium, draw, text centered, trapezium left angle=60, trapezium right angle=120, minimum height=2em, fill=blue!10]

\tikzset{fermionnoarrow/.style={draw=black},}

\def\iimg{ {\bf i}}

\definecolor{green1}{HTML}{3D792A}
\definecolor{cyan1}{HTML}{37cdaa}
\definecolor{blue1}{HTML}{5d7ac4}
\definecolor{red1}{HTML}{d0482a}
\definecolor{purple1}{HTML}{845ea8}
\definecolor{orange1}{HTML}{e07229}

\def\raj#1{({\color{red} Raj: #1})}
\def\jan#1{({\color{blue} Jan: #1})}
\def\pierpaolo#1{({\color{violet} Pierpaolo: #1})}

\title{Gravitational Quadratic-in-Spin Hamiltonian at NNNLO in the post-Newtonian framework}

\author[a]{Manoj K. Mandal,}
\author[b,a]{Pierpaolo Mastrolia,}
\author[c,d,e]{Raj Patil,}
\author[c]{Jan Steinhoff}

\newcommand{\unipd}{Dipartimento di Fisica e Astronomia, Universit\`a degli Studi di Padova,
Via Marzolo 8, I-35131 Padova, Italy.}

\newcommand{\pdinfn}{INFN, Sezione di Padova,
Via Marzolo 8, I-35131 Padova, Italy.}

\affiliation[a]{\pdinfn}
\affiliation[b]{\unipd}
\affiliation[c]{Max Planck Institute for Gravitational Physics (Albert Einstein Institute), Am M{\"u}hlenberg 1, Potsdam 14476, Germany}

\affiliation[d]{Institut f{\"u}r Physik und IRIS Adlershof, Humboldt-Universit {\"a}t zu Berlin, Zum Großen Windkanal 2, D-12489 Berlin, Germany}
\affiliation[e]{Indian Institute of Science Education and Research Bhopal, Bhopal Bypass Rd, Bhauri, Madhya Pradesh 462066, India.}

\emailAdd{\\manojkumar.mandal@pd.infn.it}
\emailAdd{\\pierpaolo.mastrolia@unipd.it}
\emailAdd{\\raj.patil@aei.mpg.de}
\emailAdd{\\jan.steinhoff@aei.mpg.de}

\abstract{
We present the result of the quadratic-in-spin interaction Hamiltonian for binary systems of rotating compact objects with generic spins, up to \NNNLO corrections within the post-Newtonian expansion. The calculation is performed by employing the effective field theory diagrammatic approach, and it involves Feynman integrals up to three loops, evaluated within the dimensional regularization scheme. 
The gauge-invariant binding energy and the scattering angle, in special kinematic regimes and spin configurations, are explicitly derived.
The results extend our earlier study on the spin-orbit interaction effects.
}

\def\NNLO{N$^2$LO }
\def\NNNLO{N$^3$LO }

\allowdisplaybreaks

\begin{document}
\addtocontents{toc}{\protect\setcounter{tocdepth}{2}}

\begin{flushright}
\begingroup\footnotesize\ttfamily
	HU-EP-22/33-RTG
\endgroup
\end{flushright}

\maketitle

\section{Introduction}

The successful detection of the gravitational waves (GW)~\cite{LIGOScientific:2016aoc} by the LIGO-Virgo-KAGRA detectors opened a new era in precision astronomy and cosmology. 
Since then, the LVK collaboration has detected around 90 GW events~\cite{LIGOScientific:2021djp}.
The primary source of these GWs are the compact binaries and 
a worldwide network of ground-based~\cite{LIGOScientific:2014pky,VIRGO:2014yos,KAGRA:2020agh,Saleem:2021iwi,LIGOScientific:2016wof,Punturo:2010zza} 
as well as space-based GW detectors~\cite{LISA} 
are coming up to explore the dynamical evolution of these compact binaries.
The compact objects in the binary may be close to maximally rotating, as seen in several recent detections \cite{Olsen:2022pin}.
Therefore, high-precision waveforms incorporating spin contributions are essential to exploit the full potential in GW astronomy.
Initial studies extending the classical techniques to include spin were done in \cite{Tulczyjew:1959,Damour:1982}, which was later extended by \cite{Tagoshi:2000zg,Faye:2006gx,Damour:2007nc,Hartung:2011te,Hartung:2013dza,Steinhoff:2008zr,Steinhoff:2009ei,Bohe:2012mr,Aoude:2022thd,FebresCordero:2022jts,Bern:2022kto,Liu:2021zxr,Kosmopoulos:2021zoq,Jakobsen:2022fcj,Jakobsen:2022zsx}. The spin effects in the post-Newtonian formalism were developed using the effective field theory approach in  \cite{Porto:2005ac,Porto:2010tr,Levi:2010zu,Levi:2020kvb,Porto:2006bt,Porto:2008tb,Levi:2008nh,Porto:2008jj,Levi:2011eq,Levi:2014sba,Levi:2015ixa,Kim:2021rfj,Levi:2020uwu,Levi:2019kgk,Levi:2015msa,Levi:2014gsa}. Another recently  developed method of using quantum scattering amplitudes involving massive particles of arbitrary spin were used to obtain classical spin corrections to the two-body effective potential \cite{Guevara:2017csg,Vines:2018gqi,Guevara:2018wpp,Chung:2018kqs,Guevara:2019fsj,Chung:2019duq,Siemonsen:2019dsu,Guevara:2020xjx,Arkani-Hamed:2019ymq}.
See \cite{Levi:2018nxp,Porto:2016pyg,Blanchet:2013haa} for recent reviews and a more comprehensive reference to literature.

In the effective field theory approach, 
the current state of the art for the post-Newtonian calculation for the conservative potential without spinning degrees of freedom is the 5PN correction computed in
\cite{Foffa:2019hrb,Blumlein:2019zku, Blumlein:2020pyo, Foffa:2020nqe, Blumlein:2021txe}. Several partial results of 6PN \cite{Blumlein:2020znm, Blumlein:2021txj} are also known. 
For the spin-orbit sector, the leading order (LO) effective potential was first computed in \cite{Porto:2005ac}. The next-to-leading order (NLO) potential in \cite{Porto:2010tr,Levi:2010zu}, and \NNLO in \cite{Levi:2015uxa}, and \NNNLO in \cite{Kim:2022pou,Mandal:2022nty}.
Similarly, 
for the quadratic in spin sector, 
the LO effective potential was computed in \cite{Porto:2005ac}; 
the NLO in \cite{Porto:2006bt,Porto:2008tb,Levi:2008nh,Porto:2008jj};
and the \NNLO in \cite{Levi:2011eq,Levi:2014sba,Levi:2015ixa}. 
Partial results for \NNNLO were reported in \cite{Kim:2021rfj,Levi:2020uwu}, 
and very recently the complete \NNNLO contribution has been reported in \cite{Kim:2022bwv}.
The computation for the cubic and higher orders in the spin variables can be found in \cite{Levi:2019kgk,Levi:2014gsa,Levi:2015msa}, 
whereas the finite size effects are described in detail in  \cite{Porto:2008jj,Levi:2014gsa,Levi:2015msa}.

In this article, following the same strategy adopted in \cite{Mandal:2022nty}, we present the complete conservative \NNNLO quadratic-in-spin interaction potential,
obtained by adopting the EFT approach proposed in~\cite{Goldberger:2004jt,Porto:2005ac} and using the computational diagrammatic techniques proposed in~\cite{Foffa:2016rgu}. 
In particular,
we begin with deriving the required Feynman rules and then the Feynman diagrams,
where we employ the Kaluza-Klein (KK) parametrization for the metric, for convenience.
The corresponding Feynman amplitudes contain tensor integrals, which are reduced to scalar integrals by means of a suitable set of projectors built out of Lorentz invariance. 
The emerging scalar integrals, up to three loop, 
are further reduced to a minimal set of independent integrals, 
dubbed master integrals (MIs), 
using the integration-by-parts (IBP) identities~\cite{Chetyrkin:1981qh,Laporta:2000dsw}.
The analytical expression of the MIs are used to obtain the analytic expression of the 
contribution of each Feynman diagram to the scattering amplitude.
Finally, the effective Lagrangian is obtained by taking 
the Fourier transform of the amplitude.
The computational procedure has been performed through in-house codes, 
in an automated manner using 
\texttt{Mathematica} routines with an 
interface to 
\texttt{QGRAF}~\cite{NOGUEIRA1993279}, generating the diagrams,
\texttt{xTensor}~\cite{xAct}, manipulating tensor algebras,
\texttt{LiteRed}~\cite{Lee:2013mka}, performing the IBP decomposition, 
inspired by several ideas implemented in \texttt{EFTofPNG}~\cite{Levi:2017kzq}.
The derived effective Lagrangian contain higher-order time derivatives 
of the position and the spin, 
which are removed by applying appropriate coordinate transformations. 
Then, the EFT Hamiltonian is derived by applying the Legendre transformation. 
However, 
it contains poles in the parameter of the dimensional regularization $\epsilon=d-3$ 
($d$ being the number of the continuous space dimensions), 
and logarithmic pieces depending on the scale of the binary system 
and they are removed following the application of appropriate canonical transformations,
which is the main result of the article.
The novel result for the quadratic-in-spin interaction Hamiltonian are further used to
compute gauge invariant observable, specifically, 
the binding energy for circular orbits with aligned spins and the scattering angle with aligned spins, and they agree with the results reported in~\cite{Kim:2022bwv}.

%\textbf{The paper.}
The paper is organised as follows. In section \ref{sec_setup}, we review
the description of the spinning binaries within the EFT formalism.  In section
\ref{sec_Routine}, we present the computation for the N$^3$LO quadratic-in-spin potential 
employing the Feynman diagrammatic approach within the EFT framework. 
Then, in section
\ref{sec_Processing_the_effective_potential}, we describe the procedure of removing the
residual divergences and logarithms to derive the EFT Hamiltonian.
We provide our main result of the quadratic-in-spin Hamiltonian up to \NNNLO in section 
\ref{sec_results}. In section \ref{sec_compputing_observables}, we compute two 
observable from the EFT Hamiltonian, namely, the binding energy of the binary system in 
circular orbits with aligned spin configuration and scattering angle for two spinning 
compact objects with aligned spins. We summarize our main results
in section \ref{sec_Conclusion}.
In appendix \ref{app_notation_and_convention}, we describe the notations used in this
article and in appendix~\ref{app_Ham}, 
we provide the Hamiltonians till \NNLO in the quadratic-in-spin sector.
\\ \\
We provide the required EFT Feynman rules 
in the ancillary file \texttt{Feynman\_Rules.m} and the analytic results of the quadratic-in-spin Hamiltonian till \NNNLO in the ancillary file \texttt{Hamiltonian.m}.
\section{EFT of spinning objects}\label{sec_setup}

In this section, 
we describe the effective action for a spinning compact object 
at the orbital scale by considering the 
degrees of freedom of the gravitational field and the degrees of freedom of the 
spinning compact objects, 
namely their center of mass and their spin. 
Then we describe the techniques of the Post-Newtonian formulation of GR in detail and briefly outline the procedure to compute the effective action.

\subsection{Action}
The effective action of the spinning compact binary can be described as 
the sum of the action of the gravitational field and the point particle effective action for each of the spinning compact objects as
\begin{align}
    S_{\text{eff}} = S_{\text{EH}} + S_{\text{pp}} \, .
\end{align}
Here $S_{\text{EH}}$ is the usual Einstein-Hilbert action and it 
expresses 
the dynamics of the gravitational field ($g_{\mu\nu}$) along with a gauge fixing term,
\begin{align}\label{eq_action_EH}
S_{\text{EH}} = -\frac{c^4}{16 \pi G_N} \int d^4x \sqrt{g} ~R[g_{\mu\nu}] + \frac{c^4}{32 \pi G_N } \int d^4x \sqrt{g} ~g_{\mu\nu}\Gamma^\mu \Gamma^\nu \, ,
\end{align}
where $\Gamma^\mu=\Gamma^{\mu}_{\rho\sigma}g^{\rho\sigma}$ (in the harmonic gauge $\Gamma^\mu = 0$), $\Gamma^{\mu}_{\rho\sigma}$ is the Christoffel symbol, $G_N$ is the Newton's constant,  $R$ is the Ricci scalar, and $g$ is the determinant of the $g_{\mu\nu}$. 

%====point particle action ======%
The spinning compact objects can be described by a point-particle effective action of each of them. 
This point-particle action can be written as an integral along a 
worldline in the following way~\cite{Levi:2015msa},
\begin{align}
\label{eq_action_pp}
S_{\text{pp}} = \sum_{a=1,2} \int d\tau \left( - m_{(a)} c \sqrt{u_{(a)}^2} -\frac{1}{2} S_{(a)\mu\nu} \Omega_{(a)}^{\mu\nu} - \frac{S_{(a)\mu\nu}u_{(a)}^\nu}{u_{(a)}^2}\frac{du_{(a)}^\mu}{d\tau} + \mathcal{L}_{(a)}^{(R)} + \mathcal{L}_{(a)}^{(R^2)} + \dots \right)\, ,
\end{align}
where $\mathcal{L}_{(a)}^{(R)}$, $\mathcal{L}_{(a)}^{(R^2)}$, etc denote Lagrangians containing nonminimal couplings at linear, quadratic, etc order in curvature specified below.
We use the Pryce, Newton, and Wigner gauge for spin supplementarity condition (SSC) given by $S_{(a)\mu\nu} (u_{(a)}^\nu + \sqrt{u_{(a)}^2} \delta^{\nu 0})\approx0$.
Here, $u_{(a)}^2=g_{\mu\nu}u_{(a)}^\mu u_{(a)}^\nu$, and $u_{(a)}^\mu$ is the four velocity, defined as $u_{(a)}^\mu=dx_{(a)}^\mu/d\tau$. 
The worldline $x_{(a)}^\mu(\tau)$ represents the center of the spinning object and is parametrized by an affine parameter $\tau$, which we are going to gauge-fix to the coordinate time $t$ by $\tau=c t$.
The $\Omega_{(a)}^{\mu\nu}$ denotes the angular velocity tensor of the spinning object 
and is defined as
\begin{align}
\Omega_{(a)}^{\mu\nu}=\Lambda^\mu_{(a)A} \frac{d\Lambda_{(a)}^{A\nu}}{d\tau} \, ,
\end{align} 
where $\Lambda^\mu_{(a)A}(\tau)$ represent the tetrads along the worldlines, connecting the body-fixed frame (denoted by upper case Latin indices) of the $a^{\text{th}}$ compact object and the general coordinate frame (denoted by Greek indices).
The $S_{(a)\mu\nu}$ are the spin tensors of the spinning objects, defined as the conjugate 
momenta to the $\Lambda^\mu_{(a)A}$ as
\begin{align}
S_{(a)\mu\nu}=-2\frac{\partial L_{\text{pp}}}{\partial \Omega_{(a)}^{\mu\nu}}\, .
\end{align}

The linear-in-curvature Lagrangian $\mathcal{L}_{(a)}^{(R)}$ consists only of the spin-induced (SI) nonminimal couplings $\mathcal{L}_{(a)\text{SI}}$~\cite{Levi:2015msa},
\begin{align}\label{eq_action_pp_R}
    \mathcal{L}_{(a)}^{(R)} \equiv \mathcal{L}_{(a)\text{SI}} &= \sum_{n=1}^{\infty} \frac{(-1)^n}{(2n)!} \frac{\Big(C_{\rm ES^{2n}}\Big)_{(a)}}{(m_{(a)} c)^{2n-1}} D_{\mu_{2n}}\cdots D_{\mu_{3}} \frac{E_{\mu_1\mu_2}}{u_{(a)}}  \Big[ S^{\mu_1}_{(a)}\cdots S^{\mu_{2n}}_{(a)} \Big]_{\rm STF}\nonumber\\
    &+\sum_{n=1}^{\infty} \frac{(-1)^n}{(2n+1)!} \frac{\Big(C_{\rm BS^{2n+1}}\Big)_{(a)}}{(m_{(a)} c)^{2n}} D_{\mu_{2n+1}}\cdots D_{\mu_{3}} \frac{B_{\mu_1\mu_2}}{u_{(a)}} \Big[ S^{\mu_1}_{(a)}\cdots S^{\mu_{2n+1}}_{(a)} \Big]_{\rm STF}
\end{align}
where STF denotes the symmetric-tracefree part (in a 3-dimensional comoving frame), which transforms irreducibly under the little group SO(3) of massive particles and hence it makes sense to construct interactions from STF building blocks.
Following the convention in~\cite{Levi:2015msa}, each Wilson coefficient is a function of the invariants, $m_{(a)}$ and the $S^2_{(a)}$.
We then expand the Wilson coefficients explicitly as a series in $S^2_{(a)}$, \begin{align}\label{eq:Wilson_expansion}
    C_{(a)} = C^{(0)}_{(a)} + C^{(2)}_{(a)} \left(\frac{S^2_{(a)} c^2}{G^2_N m^4_{(a)}}\right) + C^{(4)}_{(a)}\left(\frac{S^2_{(a)} c^2}{G^2_N m^4_{(a)}}\right)^2 + \cdots ,
\end{align}
where the $C^{(n)}_{(a)}$ are only a function of $m_{(a)}$.
Then the only contribution relevant for \NNNLO spin-squared from the spin-induced multipole interactions is given by the $\left(C^{(0)}_{\rm ES^2}\right)_{(a)}$-term,
\begin{align}\label{eq:Lnonminimal1}
    \mathcal{L}_{(a)}^{(R)} &= -\frac{1}{2m_{(a)} c} \left(C^{(0)}_{\rm ES^2}\right)_{(a)} \frac{E_{\mu\nu}}{u_{(a)}} \Big[ S_{(a)}^\mu S_{(a)}^\nu \Big]_{\rm STF} + \dots .
\end{align}

Moving on to quadratic order in curvature, due to the larger multiplicity of terms compared to $\mathcal{L}_{(a)}^{(R)}$ it makes sense to group them as $\mathcal{L}_{(a)}^{(R^2)} = \mathcal{L}_{(a)}^{(R^2, S^0)} + \mathcal{L}_{(a)}^{(R^2, S^1)} + \mathcal{L}_{(a)}^{(R^2, S^2)} + \dots$ according the power of spins in tensor contractions with curvature.
The first contribution reads
\begin{align}
    \mathcal{L}_{(a)}^{(R^2, S^0)}= \frac{1}{2} \Big(C_{\rm E^2}\Big)_{(a)} \frac{G^4_N m^5_{(a)}}{c^7} \frac{E_{\alpha\beta} E^{\alpha\beta}}{u_{(a)}^3}+ \frac{1}{2} \Big(C_{\rm \dot{E}^2}\Big)_{(a)} \frac{G^6_N m^7_{(a)}}{c^{11}} \frac{\dot{E}_{\alpha\beta} \dot{E}^{\alpha\beta}}{u_{(a)}^3} + \cdots
\end{align}
where $\dot{~} = u^{-1} u^{\mu} D_{\mu}$, and similar terms involving covariant derivatives orthogonal to $u^{\mu}$.
The first term here describes the leading-order tidal interaction entering at 5PN order, and via the expansion~\eqref{eq:Wilson_expansion} gives a spin-squared contribution reading
\begin{align}\label{eq:Lnonminimal2}
    \mathcal{L}_{(a)}^{(R^2, S^0)} &= \frac{1}{2} \left(C^{(2)}_{\rm E^2}\right)_{(a)} \frac{G_N^2 m_{(a)}}{c^5} \frac{E_{\mu\nu}E^{\mu\nu}}{u_{(a)}^3} S_{(a)}^2 + \dots .
\end{align}
The contributions from $\mathcal{L}_{(a)}^{(R^2, S^1)}$ start beyond the 5PN order and can be neglected here (since e.g. $E_{\mu\alpha}E_{\nu}^{~\alpha} S_{(a)}^{\mu\nu} \equiv 0$).
However, there is a relevant term in $\mathcal{L}_{(a)}^{(R^2, S^2)}$~\cite{Kim:2021rfj},
\begin{align}\label{eq:Lnonminimal3}
    \mathcal{L}_{(a)}^{(R^2, S^2)} &= \frac{1}{2}\left(C^{(0)}_{\rm E^2S^2}\right)_{(a)} \frac{G_N^2 m_{(a)}}{c^5} \frac{E_{\mu\alpha}E_{\nu}^{~\alpha}}{u_{(a)}^3} \Big[ S_{(a)}^\mu S_{(a)}^\nu \Big]_{\rm STF} + \dots.
\end{align}
We have now recapitulated all nonminimal spin-squared couplings relevant at \NNNLO, see \eqref{eq:Lnonminimal1}, \eqref{eq:Lnonminimal2}, and \eqref{eq:Lnonminimal3}.
We note that in order to simplify the Feynman rules, we follow the custom to drop the STF projection in \eqref{eq:Lnonminimal1} and to replace in all nonminimal couplings the Weyl tensor in $E_{\mu\nu}$, $B_{\mu\nu}$ by the Riemann tensor (being equivalent to a field redefinition).
The disadvantage of this practice is that spins in the $C_{\rm ES^2}$ coupling are not manifestly appearing in an STF combination any more, but could be brought into this form using a field redefinition.

We employ dimensional regularization for the computation of multi-loop Feynman diagrams 
and write the gravitational coupling constant in $d$ spatial dimensions as
\begin{align}
G_d = G_N \Big(\sqrt{4\pi e^{\gamma_E}} R_0\Big)^{d-3}\, ,
\end{align}
where, 
$\gamma_E$ is the Euler-Mascheroni constant, 
and $R_0$ is an arbitrary length scale.

\subsection{Post-Newtonian formulation of General Relativity}
The inspiral phase of the binary compact objects contains three widely separated
scales,
namely,
the length scale related to the single compact object ($R_s$) (Schwarzschild radius),
the orbital separation of the binary ($r$),
and the wavelength of the radiated gravitational wave ($\lambda$).
The scales have the following hierarchy 
\begin{align}
%\lambda\gg r\gg R_s\, .
R_s\ll r\ll \lambda\, .
\end{align}
As the wavelength of the radiation is much longer than the orbital separation 
and the objects are moving non-relativistically, 
we can decompose the gravitational field ($g_{\mu\nu}$)
as an expansion on the asymptotic flat space-time ($\eta_{\mu\nu}$):
$g_{\mu\nu} = \eta_{\mu\nu} + H_{\mu\nu} + \bar{h}_{\mu\nu}$, 
with two different field modes~\cite{Goldberger:2004jt}.
$H_{\mu\nu}$ is the short-distance mode (orbital) with scaling $(k_0,\textbf{k})\sim(v/r,1/r)$, mediating the gravitational interaction between the two compact objects and is known as the potential mode. 
$\bar{h}_{\mu\nu}$ is the long-distance radiation mode with scaling $(k_0,\textbf{k})\sim(v/r,v/r)$, consisting of 
the on-shell gravitons emitted from the system.

Due to the virial theorem for bound orbits $v^2 \sim 1/r$, we use $v^2 / c^2 \sim G_N M / r c^2$ as a formal dimensionless expansion, with one PN order corresponding to $1/c^2$.
Since the spin variable $S_{(a)}$ is related to the dimensionaless spin $\chi_{(a)}$ by $S_{(a)} = G m_{(a)}^2 \chi_{(a)} / c$, we rescale the spins as $S_{(a)} \rightarrow S_{(a)} / c$ in order to make the PN counting in $1/c$ manifest.

We compute the conservative potential of the binary by ignoring the radiation modes.
We further decompose the potential mode gravitons employing Kaluza-Klein (KK) 
parameterization~\cite{Kol:2007bc,Kol:2007rx}, 
where the different components of the metric 
$g_{\mu\nu}$ ($=\eta_{\mu\nu}+H_{\mu\nu}$) 
are specified by three fields, 
namely, 
a scalar $\bm{\phi}$ field,
a 3-dimensional vector $\bm{A}_i$ field,
and a 3-dimensional symmetric rank two tensor $\bm{\sigma}_{ij}$ field.
Following the KK parametrization the metric is expressed as
\begin{equation*}
g_{\mu\nu} = 
\begin{pmatrix}
e^{2\bm{\phi}/c^2} \,\,\, & -e^{2\bm{\phi}/c^2} \frac{\bm{A}_j}{c^2}\\
-e^{2\bm{\phi}/c^2} \frac{\bm{A}_i}{c^2} \,\,\,\,\,\, & -e^{-2\bm{\phi}/c^2}\bm{\gamma}_{ij}+e^{2\bm{\phi}/c^2} \frac{\bm{A}_i}{c^2}\frac{\bm{A}_j}{c^2}  
\end{pmatrix}\, ,
\end{equation*}
where, $\bm{\gamma}_{ij}=\bm{\delta}_{ij}+\bm{\sigma}_{ij}/c^2$.

The effective action for the binary at the orbital scale is then expressed by integrating out the relevant gravitational degrees of freedom from the $S_{\text{eff}}$ as
\begin{align}
\text{exp}\Big[{\iimg \int dt ~ \mathcal{L}_{\text{eff}}}\Big] = \int D\bm{\phi} D\bm{A}_i D\bm{\sigma}_{ij} ~e^{\iimg(S_{\text{EH}}+S_{\text{pp}})} \, ,
\end{align}
where, $\mathcal{L}_{\rm eff}$ signifies the effective Lagrangian and it can be further decomposed as 
\begin{equation}
 \mathcal{L}_{\rm eff} = \mathcal{K}_{\rm eff} - \mathcal{V}_{\rm eff}\, .
\end{equation}
Here, $\mathcal{K}_{\rm eff}$ denotes the kinetic term,
and $\mathcal{V}_{\rm eff}$ denotes the effective contribution due to gravitational interactions between the two compact objects.

The effective potential $\mathcal{V}_{\rm eff}$ can be demonstrated 
in terms of connected, classical, 1 particle irreducible (1PI) scattering amplitudes as
\begin{align}
\mathcal{V}_{\text{eff}}=\iimg \lim_{d\rightarrow 3} \int_\textbf{p} e^{\iimg \textbf{p}\cdot (\textbf{x}_{(1)}-\textbf{x}_{(2)})}\quad 		
\parbox{25mm}{
\begin{tikzpicture}[line width=1 pt,node distance=0.4 cm and 0.4 cm]
\coordinate[label=left: ] (v1);
\coordinate[right = of v1] (v2);
\coordinate[right = of v2] (v3);
\coordinate[right = of v3] (v4);
\coordinate[right = of v4, label=right: \tiny$(2)$] (v5);
\coordinate[below = of v1] (v6);
\coordinate[below = of v6, label=left: ] (v7);
\coordinate[right = of v7] (v8);
\coordinate[right = of v8] (v9);
\coordinate[right = of v9] (v10);
\coordinate[right = of v10, label=right: \tiny$(1)$] (v11);
\fill[black!25!white] (v8) rectangle (v4);
\draw[fermionnoarrow] (v1) -- (v5);
\draw[fermionnoarrow] (v7) -- (v11);
\end{tikzpicture}
}
\, ,
\label{eq:effective_lagrangian}
\end{align}
where $\textbf{p}$ is the transfer of momentum between the two compact objects and the diagram in the above equation represents all possible Feynman diagrams consisting of gravitons ($\bm{\phi}$, $\bm{A}_i$, and $\bm{\sigma}_{ij}$) and the two point particle denoted by the two solid black lines. The gravitons mediate the gravitational interaction between these two compact objects.

Our goal in this article is to compute the contribution of the quadratic-in-spin effective potential up to \NNNLO. 
For this purpose, we further decompose the effective kinetic and potential terms in the following way 
\begin{align}
    \mathcal{K}_{\rm eff} = \mathcal{K}_{\rm pp} + \mathcal{K}_{\rm spin}\, ,  
    \, \, \, \, \, \, \, \, \, \, \, \, \, \, \, \, 
    \mathcal{V}_{\rm eff} = \mathcal{V}_{\rm pp} + \mathcal{V}_{\rm SO} + \mathcal{V}_{\rm SS}\, ,
\end{align}
%
% $\mathcal{K}_{\rm eff} = \mathcal{K}_{\rm pp} + \mathcal{K}_{\rm spin}$ 
% and  
% $\mathcal{V}_{\rm eff} = \mathcal{V}_{\rm pp} + \mathcal{V}_{\rm spin}$,
where $\mathcal{K}_{\rm pp}$ and $\mathcal{V}_{\rm pp}$ constitutes the 
kinetic and potential contributions for the center of mass degrees of freedom of the point particles 
and 
$\mathcal{K}_{\rm spin}$, $\mathcal{V}_{\rm SO}$, and $\mathcal{V}_{\rm SS}$ constitutes the kinetic, linear in spin potential and quadratic in spin potential contributions. 

The effective kinetic terms can be expressed as
\begin{equation}
\label{eq_Kpp}
\mathcal{K}_{\rm pp} = \sum_{a=1,2} m_{(a)} 
\left[ 
\frac{1}{2}  \textbf{v}_a^2 + \frac{1}{8}  \textbf{v}_{(a)}^4 \left(\frac{1}{c^2}\right)  + \frac{1}{16}  \textbf{v}_{(a)}^6 \left(\frac{1}{c^4}\right) + \frac{5}{128}  \textbf{v}_{(a)}^8 \left(\frac{1}{c^6}\right) 
\right]
+ \mathcal{O}\left(\frac{1}{c^8}\right)\, ,
\end{equation}
\begin{align}\label{eq_KSO}
\mathcal{K}_{\rm spin} = \sum_{a=1,2} 
\Bigg\{ -\frac{1}{2} \textbf{S}_{(a)}^{ij} \bm{\Omega}_{(a)}^{ij} \left(\frac{1}{c}\right)+\textbf{S}_{(a)}^{ij}\textbf{v}_{(a)}^i\textbf{a}_{(a)}^j \left(\frac{1}{c^3}\right) &
\Bigg[
\frac{1}{2}+ \frac{3}{8}  \textbf{v}_{(a)}^2 \left(\frac{1}{c^2}\right) + \frac{5}{16}  \textbf{v}_{(a)}^4 \left(\frac{1}{c^4}\right) \nonumber\\
&+ \frac{35}{128}  \textbf{v}_{(a)}^6 \left(\frac{1}{c^6}\right)
\Bigg]
+ \mathcal{O}\left(\frac{1}{c^{11}}\right) \Bigg\}\, ,
\end{align}
and the effective potential terms can be written as follows
\begin{align}
    \mathcal{V}_{\rm pp}&= \mathcal{V}_{\rm N} + \left(\frac{1}{c^2}\right) \mathcal{V}_{\rm 1PN} + \left(\frac{1}{c^4}\right) \mathcal{V}_{\rm 2PN} + \left(\frac{1}{c^6}\right) \mathcal{V}_{\rm 3PN} + \mathcal{O}\left(\frac{1}{c^8}\right)\, ,\\
	\mathcal{V}_{\rm SO}&=  \left(\frac{1}{c^3}\right)
	\left[
	\mathcal{V}^{SO}_{\rm LO} + 
	\left(\frac{1}{c^2}\right) \mathcal{V}^{SO}_{\rm NLO} + \left(\frac{1}{c^4}\right) \mathcal{V}^{SO}_{\rm N^2LO} \right]
	+ \mathcal{O}\left(\frac{1}{c^{9}}\right) \, , \\
	\mathcal{V}_{\rm SS}&=  \left(\frac{1}{c^4}\right)
	\left[
	\mathcal{V}^{SS}_{\rm LO} + 
	\left(\frac{1}{c^2}\right) \mathcal{V}^{SS}_{\rm NLO} + \left(\frac{1}{c^4}\right) \mathcal{V}^{SS}_{\rm N^2LO} + \left(\frac{1}{c^6}\right) \mathcal{V}^{SS}_{\rm N^3LO} 
	\right]
	+ \mathcal{O}\left(\frac{1}{c^{12}}\right)\, .
\end{align}
Here, 
$\mathcal{V}_{\rm N}$ corresponds to the Newtonian potential;
$\mathcal{V}_j$ with $j={\{\rm 1PN,\, 2PN,\, 3PN\}}$ denotes the corresponding PN correction for the non-spinning part of the potential;
$\mathcal{V}^{SO}_j$ 
with $j=\{ \rm LO,\, NLO,\, N^2LO \}$ denotes the corresponding correction to the spin-orbit coupling of the binary system;
and $\mathcal{V}^{SS}_j$ 
with $j=\{ \rm LO,\, NLO,\, N^2LO,$ $\rm\, N^3LO \}$ denotes the corresponding correction to the quadratic-in-spin coupling of the binary system. Each $\mathcal{V}^{SS}_j$ is then separated with respect to their origin in the effective action as; $\rm S1S2$ and $\rm S^2$ sectors corresponding to the first three terms present given in \eqref{eq_action_pp}, $\rm ES^2$ corresponding to the action given in \eqref{eq:Lnonminimal1}, $\rm E^2$ corresponding to the action given in \eqref{eq:Lnonminimal2} and $\rm E^2S^2$ corresponding to the action given in \eqref{eq:Lnonminimal3}.
Our goal in this article is to compute the $\mathcal{V}^{SS}_{\rm N^3LO}$ employing the advanced techniques of multi-loop scattering amplitudes, as further discussed in the next sections.

\section{Computational Algorithm}
\label{sec_Routine}

\begin{table}
\centering
\begin{subtable}[H]{0.49\textwidth}
\centering
\begin{tabular}{|c|c|c|c|}
\hline
Order                & Diagrams            & Loops       & Diagrams  \\ \hline
LO                  & 1                   & 0      & 1   \\ \hline
\multirow{2}{*}{NLO} & \multirow{2}{*}{7}  & 1    & 3 \\ \cline{3-4} 
                     &                     & 0  & 4 \\ \hline
\multirow{3}{*}{\NNLO} & \multirow{3}{*}{58} & 2    & 27 \\ \cline{3-4} 
                     &                     & 1 & 24 \\ \cline{3-4} 
                     &                     & 0  & 7 \\ \hline
\multirow{4}{*}{\NNNLO} & \multirow{4}{*}{553} & 3    & 125 \\ \cline{3-4} 
                     &                     & 2 & 342 \\ \cline{3-4} 
                     &                     & 1  & 76 \\ \cline{3-4}
                     &                     & 0  & 10 \\ \hline
\end{tabular}
\caption{Spin1-Spin2 and $\rm Spin1^2$ ($\rm Spin2^2$) sector}
\end{subtable}
\begin{subtable}[H]{0.49\textwidth}
\centering
\begin{tabular}{|c|c|c|c|}
\hline
Order                & Diagrams            & Loops      & Diagrams  \\ \hline
LO                  & 1                   & 0      & 1   \\ \hline
\multirow{2}{*}{NLO} & \multirow{2}{*}{4}  & 1    & 1 \\ \cline{3-4} 
                     &                     & 0  & 3 \\ \hline
\multirow{3}{*}{\NNLO} & \multirow{3}{*}{25} & 2    & 7 \\ \cline{3-4} 
                     &                     & 1 & 12 \\ \cline{3-4} 
                     &                     & 0  & 6 \\ \hline
\multirow{4}{*}{\NNNLO} & \multirow{4}{*}{168} & 3    & 15 \\ \cline{3-4} 
                     &                     & 2 & 101 \\ \cline{3-4} 
                     &                     & 1  & 43 \\ \cline{3-4}
                     &                     & 0  & 9 \\ \hline
\end{tabular}
\caption{$\rm ES^2$ sector}
\end{subtable}

\vspace{0.5cm}
\begin{subtable}[H]{0.49\textwidth}
\centering
\begin{tabular}{|c|c|c|c|}
\hline
Order                           & Loops      & Diagrams  \\ \hline
LO                                 & 1      & 1   \\ \hline
\end{tabular}
\caption{$\rm E^2$ sector}
\end{subtable}
\begin{subtable}[H]{0.49\textwidth}
\centering
\begin{tabular}{|c|c|c|c|}
\hline
Order                          & Loops      & Diagrams  \\ \hline
LO                                 & 1      & 1   \\ \hline
\end{tabular}
\caption{$\rm E^2S^2$ sector}
\end{subtable}
\caption{Number of Feynman diagrams contributing different sectors.}
\label{tbl_no_diag_SS}
\end{table}

We employ a Feynman diagrammatic approach following equation~\eqref{eq:effective_lagrangian}
to compute the effective potential.
First, we generate all the relevant topologies contributing at different orders of $G_N$.
The virial theorem implies that contributions at N$^n$LO constitute all the terms proportional to $G_N^l$ where $l = 1, 2, ..., n + 1$, and consequently, 
we take into account all the topologies at $l-1$ loops for the contributions at specific order $l$. 
So, 
for the evaluation of the N$^3$LO quadratic-in-spin potential, 
we generate all the relevant topologies till the order $G_N^4$ (3-loop)
employing \texttt{QGRAF}~\cite{NOGUEIRA1993279}.
 There is $1$ topology at order $G_N$ (tree-level), 
$2$ topologies at order $G_N^2$ (one-loop), 
$9$ topologies at $G_N^3$ (two-loop), 
and $32$ topologies at order $G_N^4$ (three-loop).
The topologies are dressed with the KK field and 
we use the Feynman rules, 
obtained from the effective action of the PN expansion, 
to acquire all the Feynman diagrams contributing to the given order of $G_N$ and $v$,
depending on the specific perturbation order.
We provide the relevant Feynman rules after the KK parameterization in an ancillary file
\texttt{Feynman\_Rules.m} with this article.
The number of diagrams corresponding to the particular sector of quadratic-in-spin potential 
at a particular order in $1/c$ and of particular loop topology
are shown in table~\ref{tbl_no_diag_SS}\footnote{While considering the spin effects, we count only the representative Feynman diagrams, where the spin can contribute from any of the world-line graviton interaction vertex present in the diagram. 
Additionally, the diagrams, which can be obtained from the change in the label $1\leftrightarrow 2$, are not counted as separate diagrams.} 
The diagrams contributing to the non-spinning and spin-orbit sector are provided in~\cite{Mandal:2022nty}. 
\begin{figure}[H]%{R}{0.45\textwidth}%[H]
	\centering
	\begin{tikzpicture}[line width=1 pt, scale=0.4]
	\begin{scope}[shift={(-7,0)}]
	%\draw[thick, directed](0,0) ellipse (2cm and 1cm);
	\filldraw[color=gray!40, fill=gray!40, thick](0,0) rectangle (3,3);
	\draw (-1.5,0)--(4.5,0);
	\draw (-1.5,3)--(4.5,3);
	\node at (1.5,6.5) {Gravity};	
	\node at (1.5,5.3) {Diagrams};	
	\end{scope}
	\begin{scope}[shift={(0,0)}]
	\node at (0,6) {$\longleftrightarrow$};
	\node at (0,1.5) {$\equiv$};
	\end{scope}
	\begin{scope}[shift={(5,0)}]
	%\draw[thick, directed](0,0) ellipse (2cm and 1cm);
	\filldraw[color=gray!40, fill=gray!40, thick](0,1.5) circle (1.5);
	\draw (0,3)--(0,4);
	\draw (0,0)--(0,-1);	
	\node at (0,6.5) {Multi-loop};				
	\node at (0,5.3) {Diagrams};				
	\end{scope}		
	\end{tikzpicture}
	\caption{%Relation between different diagrams
	}
	\label{fig_relation_bet_diags}
\end{figure}
The generated Feynman diagrams can be understood as two-point multi-loop
Feynman diagrams with all internal lines mass-less and the external momentum, identified with the momentum transferred between two sources, as shown in figure \ref{fig_relation_bet_diags}.
We perform the tensor algebra using \texttt{xTensor} \cite{xAct} to translate the Feynman diagrams to their corresponding Feynman Amplitudes.
Consequently, we can write the generic expression of the effective potential corresponding to 
any $l$-loop Feynman graph $G$ as the following
\begin{equation}
{\cal V}_{G}^{(l)} = %(x_{(a)},S_{(a)})  = 
\underbrace{ \vphantom{\sum_{\substack{a\\ b}}} N^{\mu_1,\mu_2,\cdots}_{C~~~\nu_1,\nu_2,\cdots}
\big(x_{(a)},\cdots,S_{(a)},\cdots\big)}_{\substack{\textrm{Coefficient that depends} \\ \textrm{on orbital variables}}} ~
\underbrace{ \vphantom{\sum_{\substack{a\\ b}}} \int_p e^{\iimg p_\mu (x_{(1)}-x_{(2)})^\mu} N_{F~~~\mu_1,\mu_2,\cdots}^{\alpha_1,\alpha_2,\cdots} (p)}_\textrm{Fourier integral} ~
\underbrace{ \vphantom{\sum_{\substack{a\\ b}}} \prod_{i=1}^l \int_{k_i}  
\frac{N_{M~~~\alpha_1,\alpha_2,\cdots}^{\nu_1,\nu_2,\cdots} (k_i)}{\prod_{\sigma \in G} D_\sigma (p, k_i)}}_\textrm{Multi-loop integral}\, ,
\end{equation}
\begin{enumerate}[(i)]
\item 
$N_C$ stands for a tensor polynomial built out of the world-line coordinates ($x_{(a)}^\mu$), 
the spin tensor ($S_{(a)\mu\nu}$), 
and their higher-order time derivatives,
\item $p$ denotes the transfer of the momentum between the sources (Fourier momentum), 
\item $N_F$ is the tensor polynomial depending on $p$,
\item $k_i$ denotes the loop momentum, 
\item $D_\sigma$ is the set of denominators corresponding to the graph  $G$,
\item $N_M$ is a tensor polynomial depending on external momenta $p$ and loop momenta $k_i$.
\end{enumerate}
We employ a set of projectors to reduce the multi-loop tensor integrals to scalar integrals.
The projectors are built with the Lorentz invariant external momentum($p$) and the background metric by exploiting Lorentz invariance.
With the application of the projectors, the numerator ($N_M$) of the multi-loop integral translates to
\begin{equation}
    N_{M~~~\alpha_1,\alpha_2,\cdots}^{\nu_1,\nu_2,\cdots} (p,k_i)
    \longrightarrow 
    \tilde{N}_{M} (p,k_i) \tilde{N}_{F~~~\alpha_1,\alpha_2,\cdots}^{\nu_1,\nu_2,\cdots}(p) \, ,
\end{equation}
where,
$\tilde{N}_{M}$ is a polynomial, which depends on the scalar products built out of 
external momentum ($p$) and the loop momentums ($k_i$).
Specifically, we use projectors up to rank 6 to compute the \NNNLO quadratic-in-spin effective potential.
Following this procedure, we obtain many scalar integrals, which are not all independent.
There are linear relations between these scalar Feynman integrals, 
thanks to dimensional regularization, 
known as Integration-By-Parts (IBP) relations~\cite{Chetyrkin:1981qh}.
These IBP identities can be generated and solved algorithmically and automatically, thanks to several publicly available automatic IBP solvers~\cite{Lee:2013mka,vonManteuffel:2012np,Maierhofer:2017gsa}.
Specifically, we use \texttt{LiteRED} \cite{Lee:2013mka} to build these IBP identities and solve them,
thus obtaining a smaller set of independent scalar integrals, known as Master integrals (MIs).
For the entire computation of \NNNLO quadratic-in-spin effective potential,
we obtain 1 MI at one loop, 
2 MIs at two-loop, 
and 3 MIs at three-loop.
The MIs till three-loop are already known in the literature
and have closed analytic expressions in $d$ dimension.
The expressions of these MIs are provided
in~appendix B of \cite{Mandal:2022nty}.

With the evaluation of the multi-loop integrals,
we apply the Fourier transform to the tensor polynomials, which have the following generic form
\begin{equation}
%    {\cal M}_F = 
    \int_p e^{p_\mu (x_{(1)}-x_{(2)})^\mu} N_{F~~~\mu_1,\mu_2,\cdots}^{\alpha_1,\alpha_2,\cdots} (p) \tilde{N}_{F~~~\alpha_1,\alpha_2,\cdots}^{\nu_1,\nu_2,\cdots}(p) ~ f(p)\, .
\end{equation}
Here, $N_F$ and $\tilde{N}_F$ are the tensor polynomials depending on 
the Fourier momenta $p$ and $f(p)$ is a function of the 
external momentum obtained from the evaluation of the multi-loop integrals.
We require Fourier integrals till rank 8.
%They can be obtained by performing iterative differentiation of the expression in section \ref{sec_fourierints} in the appendix.
After the Fourier transformation, 
we expand the expression as a Laurent series in $\epsilon$ around $d=3$ 
and obtain the effective potential, 
which depends on the orbital variables, namely, $\textbf{x}_{(a)}$, $\textbf{S}_{(a)}$, and their higher-order time derivatives.

We perform the generation of the relevant Feynman diagrams, 
translation of them to their corresponding multi-loop integrands, 
application of IBP reduction, 
and the Fourier transformation via our in-house code in an automated manner, 
thus obtaining the full effective potential. 
The different steps of this procedure have been explained using a flow chart in~\cite{Mandal:2022nty}.

\section{Processing the effective Lagrangian} 
\label{sec_Processing_the_effective_potential}
The conservative quadratic-in-spin potential at N$^3$LO contains higher-order 
time derivatives of the
position ($\textbf{a}_{(a)}$, $\dot{\textbf{a}}_{(a)}$, $\ddot{\textbf{a}}_{(a)}$,$\cdots$) 
and the spin ($\dot{\textbf{S}}_{(a)}$, $\ddot{\textbf{S}}_{(a)}$,$\cdots$) of the compact objects.
Specifically, we obtain the 6th-order time derivative in the position and the 5th-order time derivative in the spin. 
Besides, the effective potential in the non-spinning sector at 3PN, 
the N$^3$LO quadratic-in-spin potential also contains poles 
in the dimensional regularization parameter
$\epsilon$ and the logarithmic terms of the form log($r \over R_0$). 
We must eliminate higher-order time derivatives and the divergences to obtain the EFT Hamiltonian.
All the logarithms in this computation result from a poor choice of coordinates, 
as there are no tail effects present either at the 3PN non-spinning sector or \NNNLO quadratic-in-spin sector. So, we can define a set of suitable coordinate transformations to eliminate these contributions.

First, we describe the elimination procedure of the higher-order time derivatives.
We then demonstrate the appropriate coordinate transformations required to remove the divergent and the logarithmic terms, thus obtaining the EFT Hamiltonian free of them.

\subsection{Elimination of higher-order time derivatives}
\label{sec_Elimination_of_higher_order_time_derivatives}

Here, we discuss the procedure for the removal of all the higher-order time derivatives, 
by defining appropriate 
coordinate transformations~\cite{Levi:2014sba,Schafer:1984mr,Damour:1990jh,Damour:1985mt,Barker:1980spx}.
Under a small arbitrary shift in the coordinate $\textbf{x}_{(a)} \rightarrow \textbf{x}_{(a)} + \delta \textbf{x}_{(a)}$, 
the change in the Lagrangian is 
\begin{align}
    \delta \mathcal{L} = \left( \frac{\delta \mathcal{L}}{\delta \textbf{x}_{(a)}^i} \right) \delta \textbf{x}_{(a)}^i + \frac{1}{2}\left( \frac{\delta^2 \mathcal{L}}{\delta \textbf{x}_{(a)}^i \delta \textbf{x}_{(a)}^j} \right) \delta \textbf{x}_{(a)}^i \delta \textbf{x}_{(a)}^j+\mathcal{O}\left(\delta \textbf{x}_{(a)}^3\right) \, .
\label{eq:change_in_L}    
\end{align}
We can choose the small arbitrary shift $\delta \textbf{x}_{(a)}$ in such a way that $\mathcal{L}+\delta \mathcal{L}$ is free of the $\textbf{a}_{(a)}$, assuming that the equation~\eqref{eq:change_in_L} is linear in $\textbf{a}_{(a)}$ at LO.
We can also apply the total time derivative on the $\delta \textbf{x}_{(a)}$ to modify the equations in such a way that terms involving higher-order time derivatives of $\textbf{a}_{(a)}$ are removed.
Following the same strategy, we can define a small arbitrary transformation simultaneously to the rotation matrix $\bm{\Lambda}_{(a)}^{ij} \rightarrow \bm{\Lambda}_{(a)}^{ij} + \delta\bm{\Lambda}_{(a)}^{ij}$ and the spin $\bm{S}_{(a)}^{ij} \rightarrow \bm{S}_{(a)}^{ij} + \delta\bm{S}_{(a)}^{ij}$ to eliminate the higher-order time derivatives in spin.
A generic rotation can be expressed in terms of a matrix exponential $e^{\bm{\omega}_{(a)}}$ such that the transformation of the rotation matrix and the spin can be described as $\bm{\Lambda}_{(a)} \rightarrow \bm{\Lambda}_{(a)} e^{\bm{\omega}_{(a)}}$ and $\bm{S}_{(a)} \rightarrow e^{-\bm{\omega}_{(a)}} \bm{S}_{(a)} e^{\bm{\omega}_{(a)}}$, correspondingly. Consequently, the small arbitrary shift in the rotation matrix becomes
\begin{align}
\delta\bm{\Lambda}_{(a)}^{ij} = \bm{\Lambda}_{(a)}^{ik} \bm{\omega}_{(a)}^{kj} 
+ \mathcal{O}\left(\bm{\omega}_{(a)}^2\right) \ ,
\end{align}
where the $\bm{\omega}_{(a)}^{kj}$ denotes the antisymmetric generator of the rotation matrix,
and similarly the shift in spin becomes
\begin{align}
\delta\bm{S}_{(a)}^{ij} = 2 \bm{S}_{(a)}^{k[i} \bm{\omega}_{(a)}^{j]k} 
+ \mathcal{O}\left(\bm{\omega}_{(a)}^2\right) \ .
\end{align}
Following these transformations, 
the 
change in the Lagrangian\footnote{Here $V\equiv -\left(\mathcal{L}-\left(-\frac{1}{2} \textbf{S}_{(a)}^{ij} \bm{\Omega}_{(a)}^{ij}\right)\right)$.} becomes
\begin{align}
    \delta \mathcal{L} = &%(\textit{EOM for spin})^{ij}~\bm{\omega}_{(a)}^{ij} + \mathcal{O}\left(\bm{\omega}_{(a)}^2\right)
    - \left(\frac{1}{c}\right) \frac{1}{2} \dot{\bm{S}}_{(a)}^{ij} \bm{\omega}_{(a)}^{ij}  - \left(\frac{\delta V }{\delta \bm{S}_{(a)}^{ij} }\right) \delta\bm{S}_{(a)}^{ij}  + \mathcal{O}\left(\bm{\omega}_{(a)}^2, \delta\bm{S}_{(a)}^2\right) \, .
\end{align}
So, assuming the equation of motion is linear in $\dot{\textbf{S}}_{(a)}$ at LO, 
we can build the $\bm{\omega}_{(a)}^{ij}$ in such a way that $\mathcal{L}+\delta \mathcal{L}$
is free of $\dot{\textbf{S}}_{(a)}$. 
We can also apply the total time derivative on the equation of motion to define the modified equations in terms of higher-order time derivatives of $\dot{\textbf{S}}_{(a)}$, 
thus removing them from the new effective Lagrangian.

We use this approach iteratively for the elimination of the higher-order time derivatives from the \NNNLO quadratic-in-spin potential. 
Specifically, we perform 5 iterations, where
\begin{enumerate}
    \item we remove the terms with $\textbf{a}_{(a)}$ and its higher-order time derivatives from the LO and NLO spin-orbit potentials,
    \item we remove  $\dot{\textbf{S}}_{(a)}$ and its higher-order time derivatives from the NLO spin-orbit potential,
    \item we remove the $\textbf{a}_{(a)}$ and its higher-order time derivatives from non-spinning 2PN and 3PN  potentials as well as LO and NLO quadratic-in-spin potentials,
    \item we remove  $\dot{\textbf{S}}_{(a)}$ and its higher-order time derivatives from the NLO quadratic-in-spin potential,
    \item we remove the $\textbf{a}_{(a)}$, $\dot{\textbf{S}}_{(a)}$ and their higher-order time derivatives from the \NNLO spin-orbit potential as well as \NNLO and \NNNLO quadratic-in-spin potential.
\end{enumerate}
After each iteration, we obtain a new Lagrangian and compute the equation of motion again to be used in the next iteration.
One can check that at each step contributions cubic in $\delta\textbf{x}_{(a)}$ and quadratic in $\bm{\omega}_{(a)}$ are negligible (higher order in spin or the PN approximation), in contrast to an (incorrect) insertion of all higher-order time derivates in a single step.
Following these steps, we obtain the effective Lagrangian, which depends on the position, velocity, and spin only.

\subsection{Computation of the EFT Hamiltonians}
We derive the EFT Hamiltonian by applying the Legendre transformation on the effective Lagrangian
\begin{align}\label{eq_ham_def}
\mathcal{H}(\textbf{x},\textbf{p},\textbf{S})= \sum_{a=1,2}\textbf{p}_{(a)}^i \dot{\textbf{x}}_{(a)}^i - \mathcal{L}(\textbf{x},\dot{\textbf{x}},\textbf{S}) \ ,
\end{align}
where $\textbf{p}^i$ denotes the canonical momenta and it is defined as
\begin{align}\label{eq_conjmom_def}
\textbf{p}_{(a)}^i=\frac{\partial \mathcal{L}(\textbf{x},\dot{\textbf{x}},\textbf{S})}{\partial \dot{\textbf{x}}_{(a)}^i}\, .
\end{align}
We express $\dot{\textbf{x}}_{(a)}^i$ in terms of $\textbf{p}_{(a)}^i$ 
by inverting this relation in every order of $1/c$.
Exploiting this equation, 
we obtain a relation between $\dot{\textbf{x}}_{(a)}^i$ and 
$\textbf{p}_{(a)}^i$ and using it in the equation \eqref{eq_ham_def} 
we obtain the the required Hamiltonian $\mathcal{H}(\textbf{x},\textbf{p},\textbf{S})$.

\subsection{Removal of the poles and logarithms}

Both the Hamiltonian and the Lagrangian obtained in the previous step contain divergent pieces and logarithmic terms, which can be eliminated following appropriate coordinate transformations.
In the Lagrangian description, we can employ a set of total derivative terms with arbitrary coefficients, which are determined by demanding the cancellation of the divergent pieces during the elimination of the higher-order time derivatives as described in section~\ref{sec_Elimination_of_higher_order_time_derivatives}.
In the Hamiltonian description, we can utilize the properties of the canonical transformation
\footnote{The Poisson bracket is defined as
\begin{equation}
    \{A,B\} = \sum_{i=1}^N\left(\frac{\partial A}{\partial r_i}\frac{\partial B}{\partial p_i}-\frac{\partial A}{\partial p_i}\frac{\partial B}{\partial r_i}\right)
\end{equation}
}
to remove the poles and the logarithmic terms, namely 
\begin{align}
\mathcal{H}'=\mathcal{H}+\{\mathcal{H},\mathcal{G}\} \ ,
\label{eq:Hamiltonian_Cannonical_Transformation}
\end{align}
where $\mathcal{G}$ represents the infinitesimal generator of the arbitrary canonical transformation and $\mathcal{H}'$ is free of the divergences and logarithmic terms.

Here, we follow the procedure of removal of the poles and the logarithms in the Hamiltonian description, where we derive all the necessary suitable
canonical transformations.
Specifically, we construct an ansatz with several arbitrary coefficients for the infinitesimal generator $\mathcal{G}$ and build a system of linear equations in terms of those unknown coefficients by requiring the elimination of the divergent pieces in $\mathcal{H}'$. 
The solutions of this system of equations provide a set of values for the arbitrary coefficients, thus obtaining the final effective Hamiltonian free of poles and logarithmic terms.
In the next sections, we illustrate the strategies for the elimination of the poles and logarithmic terms from the 3PN non-spinning sector and the quadratic-in-spin \NNNLO sector.

\subsubsection{3PN non-spinning sector}
In the non-spinning sector the 3PN corrections has been known for a long time~\cite{Foffa:2011ub}. 
Following~\cite{Foffa:2011ub}, we pursue the same procedure, by adding a total derivative term with the complete 3PN Lagrangian, to remove the divergent pieces.
\begin{align}
\mathcal{L}_{\text{TD}}=\left(\frac{1}{c^6}\right)\frac{1}{\epsilon}\frac{d}{dt}\left[\frac{G_N^3}{r}\Big(c_1 \left(\textbf{v}_{(1)} \cdot \textbf{n}\right) + c_2 \left(\textbf{v}_{(2)} \cdot \textbf{n}\right)\Big)\right]\, ,
\end{align}
with,
\begin{align}
c_1= \frac{1}{3} \left(4  m_{(1)}^3 m_{(2)}- m_{(1)} m_{(2)}^3\right)\quad\quad,\quad\quad c_2= \frac{1}{3} \left( m_{(1)}^3 m_{(2)}-4  m_{(1)} m_{(2)}^3\right)\, .
\end{align}
We start adopting the procedures for the removal of $\textbf{a}^i$ and $\dot{\textbf{S}}^{ij}$ and their higher-order time derivatives as mentioned in Sec.~\ref{sec_Elimination_of_higher_order_time_derivatives}
with the 3PN Lagrangian along with the total derivative term.
Ultimately, 
the divergent pieces are eliminated, 
thus deriving a finite Lagrangian as well as free of any higher-order time derivatives till 3PN.
We build the Hamiltonian by applying the Legendre transformation on this Lagrangian and the 
Hamiltonain still contains the logarithmic terms.
So, 
we construct the ansatz needed to define an appropriate canonical transformation in the Hamiltonian description to remove these logarithmic terms.
We build the following generic ansatz for the arbitrary infinitesimal generator as,
\begin{align}
\mathcal{G}_{\text{3PN}}=&\left(\frac{1}{c^6}\right)\frac{G_N^3}{r^2}\left( g_1  \frac{1}{m_{(1)}} \left(\textbf{p}_{(1)} \cdot \textbf{n}\right) + g_2 \frac{1}{m_{(2)}} \left(\textbf{p}_{(2)} \cdot \textbf{n}\right)  \right)\, ,
\end{align}
with,
\begin{align}
\label{eq:can_trans_unknown_coeff}
g_n \equiv g_{Ln} \log\left(\frac{r}{R_0}\right)\, \, , \, \, \text{for} \,\, n=1,2 \, .
\end{align}
Using this generator, 
we obtain the canonical transformation following Eq.~\ref{eq:Hamiltonian_Cannonical_Transformation}. 
We construct a system of linear equations in the unknown coefficients, defined
in Eq.~\eqref{eq:can_trans_unknown_coeff} by demanding the cancellation of the 
logarithmic pieces. We solve these set of equations to obtain
\begin{align}
g_{L1}= m_{(1)} m_{(2)}^3-4 m_{(1)}^3 m_{(2)} \quad\quad,\quad\quad g_{L2}= 4 m_{(1)} m_{(2)}^3-m_{(1)}^3 m_{(2)}\, .
\end{align}

\subsubsection{Quadratic-in-spin \NNNLO sector}
In the quadratic-in-spin sector at \NNNLO, 
we eliminate both the divergent and the logarithmic pieces by finding an appropriate canonical transformation in the Hamiltonian description. 
%\mkm{we need the defintion of the Hamiltonian in quadratic-in-spin sector at \NNNLO}.
For this purpose, we construct the following ansatz for the infinitesimal
generator stimulated by~\cite{Levi:2014sba},
\begin{align}
    \mathcal{G}_{\rm SS-N^3LO}=\mathcal{G}_{\rm S1S2-N^3LO}+\mathcal{G}_{\rm S^2-N^3LO}
\end{align}
where,
\begin{align}
\label{eq:can_trans_ansatz_s1s2}
\mathcal{G}_{\rm S1S2-N^3LO}=\left(\frac{1}{c^{10}}\right)\frac{G_N^3}{r^4}&\Bigg\{ 
\Big(\textbf{S}_{(1)}^{ij} \textbf{S}_{(2)}^{ij}\Big) \left(g_{3} \frac{1}{m_{(1)}} \left(\textbf{p}_{(1)} \cdot \textbf{n}\right) + g_{4} \frac{1}{m_{(2)}} \left(\textbf{p}_{(2)} \cdot \textbf{n}\right)\right)\nonumber\\
&\quad +\Big(\textbf{S}_{(1)}^{ik} \textbf{S}_{(2)}^{jk} \textbf{n}^i\textbf{n}^j\Big) \left(g_{5} \frac{1}{m_{(1)}} \left(\textbf{p}_{(1)} \cdot \textbf{n}\right) + g_{6} \frac{1}{m_{(2)}} \left(\textbf{p}_{(2)} \cdot \textbf{n}\right)\right)\nonumber\\
&\quad + g_{7} \Big(\textbf{S}_{(1)}^{ik} \textbf{S}_{(2)}^{jk} \textbf{p}_{(1)}^i\textbf{n}^j\Big) + g_{8} \Big(\textbf{S}_{(1)}^{ik} \textbf{S}_{(2)}^{jk} \textbf{n}^i \textbf{p}_{(1)}^j\Big)\nonumber\\
&\quad + g_{9} \Big(\textbf{S}_{(1)}^{ik} \textbf{S}_{(2)}^{jk} \textbf{p}_{(2)}^i\textbf{n}^j\Big) + g_{10} \Big(\textbf{S}_{(1)}^{ik} \textbf{S}_{(2)}^{jk} \textbf{n}^i \textbf{p}_{(2)}^j\Big)
 \Bigg\}\, ,
\end{align}
\begin{align}
\label{eq:can_trans_ansatz_s2}
\mathcal{G}_{\rm S^2-N^3LO}=\left(\frac{1}{c^{10}}\right)\frac{G_N^3}{r^4}&\Bigg\{ 
\Big(\textbf{S}_{(1)}^{ij} \textbf{S}_{(1)}^{ij}\Big) \left(g_{11} \frac{1}{m_{(1)}} \left(\textbf{p}_{(1)} \cdot \textbf{n}\right) + g_{12} \frac{1}{m_{(2)}} \left(\textbf{p}_{(2)} \cdot \textbf{n}\right)\right)\nonumber\\
&\quad +\Big(\textbf{S}_{(1)}^{ik} \textbf{S}_{(1)}^{jk} \textbf{n}^i\textbf{n}^j\Big) \left(g_{13} \frac{1}{m_{(1)}} \left(\textbf{p}_{(1)} \cdot \textbf{n}\right) + g_{14} \frac{1}{m_{(2)}} \left(\textbf{p}_{(2)} \cdot \textbf{n}\right)\right)\nonumber\\
&\quad + g_{15} \Big(\textbf{S}_{(1)}^{ik} \textbf{S}_{(1)}^{jk} \textbf{p}_{(1)}^i\textbf{n}^j\Big) + g_{16} \Big(\textbf{S}_{(1)}^{ik} \textbf{S}_{(1)}^{jk} \textbf{p}_{(2)}^i\textbf{n}^j\Big) 
 \Bigg\}\, +(1\leftrightarrow 2) \, .
\end{align}
The $g_n$s are the unknown coefficients and defined as
\begin{align}
\label{eq:can_trans_unknown_coeff_spin2}
g_n \equiv g_{\epsilon n} \frac{1}{\epsilon} + g_{Ln} \log\left(\frac{r}{R_0}\right)\, , \, \, \text{for} \,\, n=3,4,\cdots 16 \, .
\end{align}
Following Eq.~\eqref{eq:Hamiltonian_Cannonical_Transformation},
we use the ansatz for the infinitesimal generator defined in 
%Eq.~\eqref{eq: can_trans_ansatz_s1s2}, Eq.~\eqref{eq: can_trans_ansatz_s2}
Eq.~(\ref{eq:can_trans_ansatz_s1s2},~\ref{eq:can_trans_ansatz_s2})
to obtain the canonical transformation
and prepare a system of linear equation in terms of the unknown coefficients in Eq.~\eqref{eq:can_trans_unknown_coeff_spin2} by requiring the cancellation of the divergent as well as the logarithmic pieces. 
Next, we solve them to determine the arbitrary coefficients as following
\begin{align}
&g_{\epsilon 3} = \frac{1}{10} \left(5 m_{(2)}^2-61 m_{(1)}^2\right) \, , && g_{\epsilon 4}= \frac{1}{10} \left(61 m_{(2)}^2-5 m_{(1)}^2\right)\nonumber\, ,\\
&g_{\epsilon 5}= \frac{1}{2} \left(61 m_{(1)}^2-5 m_{(2)}^2\right) \, , && g_{\epsilon 6}= \frac{1}{2} \left(5 m_{(1)}^2-61 m_{(2)}^2\right)\nonumber\, ,\\
&g_{\epsilon 7}= -\frac{61 m_{(1)}^2-5 m_{(2)}^2}{10 m_{(1)}} \, , && g_{\epsilon 8}= -\frac{61 m_{(1)}^2-5 m_{(2)}^2}{10 m_{(1)}}\nonumber\, ,\\
&g_{\epsilon 9}=-\frac{5 m_{(1)}^2-61 m_{(2)}^2}{10 m_{(2)}} \, , && g_{\epsilon 10}= -\frac{5 m_{(1)}^2-61 m_{(2)}^2}{10 m_{(2)}}\nonumber\, ,\\
&g_{L3}= \frac{3}{10} \left(61 m_{(1)}^2-5 m_{(2)}^2\right) \, , &&g_{L4}= \frac{3}{10} \left(5 m_{(1)}^2-61 m_{(2)}^2\right)\nonumber\, ,\\
&g_{L5}= \frac{1}{2} (-3) \left(61 m_{(1)}^2-5 m_{(2)}^2\right)\, , &&g_{L6}= \frac{1}{2} (-3) \left(5 m_{(1)}^2-61 m_{(2)}^2\right)\nonumber\, ,\\
&g_{L7}= \frac{3 \left(61 m_{(1)}^2-5 m_{(2)}^2\right)}{10 m_{(1)}}\, , &&g_{L8}= \frac{3 \left(61 m_{(1)}^2-5 m_{(2)}^2\right)}{10 m_{(1)}}\nonumber\, ,\\
&g_{L9}= \frac{3 \left(5 m_{(1)}^2-61 m_{(2)}^2\right)}{10 m_{(2)}}\, , &&g_{L10}= \frac{3 \left(5 m_{(1)}^2-61 m_{(2)}^2\right)}{10 m_{(2)}}\nonumber\, ,
\end{align}
\begin{align}
&g_{\epsilon 11} = \frac{1}{70} \left(2 m_1 m_2-159 \left(C^{(0)}_{\rm ES^2}\right)_{(1)} m_1 m_2\right)  \, , && g_{\epsilon 12}= -\frac{\left(C^{(0)}_{\rm ES^2}\right)_{(1)}(-175  m_2^3+16 m_1^2 m_2)+30 m_1^2 m_2}{70 m_1}\nonumber\, ,\\
&g_{\epsilon 13}= \frac{1}{14} \left(159 \left(C^{(0)}_{\rm ES^2}\right)_{(1)} m_1 m_2-2 m_1 m_2\right) \, , && g_{\epsilon 14}= -\frac{\left(C^{(0)}_{\rm ES^2}\right)_{(1)}(175  m_2^3-16 m_1^2 m_2)-30 m_1^2 m_2}{14 m_1}\nonumber\, ,\\
&g_{\epsilon 15}= \frac{1}{35} \left(2 m_2-159 \left(C^{(0)}_{\rm ES^2}\right)_{(1)} m_2\right) \, , && g_{\epsilon 16}= -\frac{\left(C^{(0)}_{\rm ES^2}\right)_{(1)}(16  m_1^2-175  m_2^2)+30 m_1^2}{35 m_1}\nonumber\, ,\\
&g_{L11}= \frac{3}{70} \left(159 \left(C^{(0)}_{\rm ES^2}\right)_{(1)} m_1 m_2-2 m_1 m_2\right) \, , &&g_{L12}= \frac{3 \left(\left(C^{(0)}_{\rm ES^2}\right)_{(1)}(-175  m_2^3+16  m_1^2 m_2)+30 m_1^2 m_2\right)}{70 m_1}\nonumber\, ,\\
&g_{L13}=\frac{(-3)}{14}  \left(159 \left(C^{(0)}_{\rm ES^2}\right)_{(1)} m_1 m_2-2 m_1 m_2\right)\, , &&g_{L14}= -\frac{3 \left(\left(C^{(0)}_{\rm ES^2}\right)_{(1)}(-175 m_2^3+16 m_1^2 m_2)+30 m_1^2 m_2\right)}{14 m_1}\nonumber\, ,\\
&g_{L15}= \frac{3}{35} \left(159 \left(C^{(0)}_{\rm ES^2}\right)_{(1)} m_2-2 m_2\right)\, , &&g_{L16}= \frac{3 \left(\left(C^{(0)}_{\rm ES^2}\right)_{(1)}(16  m_1^2-175 m_2^2)+30 m_1^2\right)}{35 m_1}\nonumber\, ,
\end{align}
 Ultimately, following these steps, we derive the full effective Hamiltonian, 
 which is free of the divergent and logarithmic pieces,
 in the quadratic-in-spin sector till \NNNLO.

\section{Results}\label{sec_results}
In this section, we illustrate the results for the effective Hamiltonian till \NNNLO.
For the convenience, we introduce a set of dimensionless variables, and write the Hamiltonian in terms of these variables, thus obtaining a compact form.
We define the total mass $M=m_{(1)}+m_{(2)}$, and the reduced mass of the two body system $\mu=m_{(1)}m_{(2)}/M$, the mass ratio $q=m_{(1)}/m_{(2)}$, the symmetric mass ratio $\nu=\mu/M$, and the antisymmetric mass ratio $\delta=(m_{(1)}-m_{(2)})/M$. The relations between these dimensionless variables can be expressed as
\begin{align}
\nu=\frac{m_{(1)} m_{(2)}}{M^2}=\frac{\mu}{M}=\frac{q}{(1+q)^2} = \frac{(1-\delta^2)}{4}\, .
\end{align}
Furthermore, 
we choose the center of mass (COM) frame of reference and show the Hamiltonian in the COM frame, for convenience.
The momentum in the COM frame is defined as $\textbf{p} \equiv \textbf{p}_{(1)}=-\textbf{p}_{(2)}$, and the orbital angular momentum 
can be expressed as $\textbf{L}=(\textbf{r}\times \textbf{p})$.
Consequently, we obtain $p^2 = \widetilde{p}_r^2 + \widetilde{L}^2 / \widetilde{r}^2$, where $\widetilde{p}_r=\textbf{p}\cdot \textbf{n}$, $p\equiv |\textbf{p}|$ and $L\equiv |\textbf{L}|$. Specifically, we use the following definitions in terms of dimensionless parameters
\begin{align}
\widetilde{\textbf{p}}=\frac{\textbf{p}}{\mu c} \quad\, ,\quad\quad \widetilde{\textbf{r}}=\frac{\textbf{r}~c^2}{G_N M} \quad\, ,\quad\quad \widetilde{\textbf{L}}=\frac{\textbf{L}~c}{G_N M \mu} \quad\, ,\quad\quad \widetilde{\textbf{S}}_{(a)}=\frac{\textbf{S}_{(a)}}{G_N M \mu} \quad\, ,\quad\quad \widetilde{\mathcal{H}}=\frac{\mathcal{H}}{\mu c^2}\, .
\end{align}
The total EFT Hamiltonian contributing towards the quadratic-in-spin sector till \NNNLO is written as 
\begin{align}
\widetilde{\mathcal{H}} = \widetilde{\mathcal{H}}_{\rm pp} + \widetilde{\mathcal{H}}_{\rm SO} + \widetilde{\mathcal{H}}_{\rm SS}\ ,
\end{align}
where, 
\begin{align}\label{eq_Ham_pp}
\widetilde{\mathcal{H}}_{\rm pp} &= \widetilde{\mathcal{H}}_{\text{0PN}}+\left(\frac{1}{c^2}\right)\widetilde{\mathcal{H}}_{\text{1PN}}+\left(\frac{1}{c^4}\right)\widetilde{\mathcal{H}}_{\text{2PN}}+\left(\frac{1}{c^6}\right)\widetilde{\mathcal{H}}_{\text{3PN}}+\mathcal{O}\left(\frac{1}{c^8}\right) \ , \\
\label{eq_Ham_SO}
\widetilde{\mathcal{H}}_{\rm SO} &= \left(\frac{1}{c^3}\right)\widetilde{\mathcal{H}}^{\rm SO}_{\text{LO}}+\left(\frac{1}{c^5}\right)\widetilde{\mathcal{H}}^{\rm SO}_{\text{NLO}}+\left(\frac{1}{c^7}\right)\widetilde{\mathcal{H}}^{\rm SO}_{\rm N^2LO}+\mathcal{O}\left(\frac{1}{c^{9}}\right)\, ,\\
\label{eq_Ham_S1S2}
\widetilde{\mathcal{H}}_{\rm SS} &= \left(\frac{1}{c^4}\right)\left(
\widetilde{\mathcal{H}}^{\rm S1S2}_{\text{LO}}+\widetilde{\mathcal{H}}^{\rm S^2}_{\text{LO}}+\widetilde{\mathcal{H}}^{\rm ES^2}_{\text{LO}}
\right)
+\left(\frac{1}{c^6}\right)\left(
\widetilde{\mathcal{H}}^{\rm S1S2}_{\text{NLO}}+\widetilde{\mathcal{H}}^{\rm S^2}_{\text{NLO}}+\widetilde{\mathcal{H}}^{\rm ES^2}_{\text{NLO}}
\right)\nonumber\\
&+\left(\frac{1}{c^8}\right)\left(
\widetilde{\mathcal{H}}^{\rm S1S2}_{\text{\NNLO}}+\widetilde{\mathcal{H}}^{\rm S^2}_{\text{\NNLO}}+\widetilde{\mathcal{H}}^{\rm ES^2}_{\text{\NNLO}}
\right)\nonumber\\
&+\left(\frac{1}{c^{10}}\right)\left(
\widetilde{\mathcal{H}}^{\rm S1S2}_{\text{\NNNLO}}+\widetilde{\mathcal{H}}^{\rm S^2}_{\text{\NNNLO}}+\widetilde{\mathcal{H}}^{\rm ES^2}_{\text{\NNNLO}}+\widetilde{\mathcal{H}}^{\rm E^2S^2}_{\rm LO}+\widetilde{\mathcal{H}}^{\rm E^2}_{\rm LO}
\right)+\mathcal{O}\left(\frac{1}{c^{12}}\right)\,
\end{align}
In the non-spinning part, the Hamiltonian till 3PN is known in the literature~\cite{Foffa:2011ub} 
and in the spin-orbit sector the Hamiltonian is known till \NNLO\cite{Levi:2015uxa}. 
We present the novel result of the $\tilde{\mathcal{H}}^{\rm SS}_{\rm N^3LO}$ in the COM frame
following an EFT approach in the following as 
%\raj{Can we make the Hamiltonian explicitly symmetric in $S1\leftrightarrow S2$ and $q\leftrightarrow 1/q$.}
\begin{align}
\widetilde{\mathcal{H}}^{\rm S1S2}_{\rm N^3LO}&= 
\left(\widetilde{\textbf{S}}_{(1)}\cdot\widetilde{\textbf{S}}_{(2)}\right)\nonumber\\
&~~~~~~~~~~ \Bigg\{
\widetilde{L}^6 \left(-\frac{5 \nu ^4}{64 \widetilde{r}^9}+\frac{7 \nu ^3}{32 \widetilde{r}^9}-\frac{45 \nu ^2}{16 \widetilde{r}^9}+\frac{5 \nu }{8 \widetilde{r}^9}\right)\nonumber\\
&~~~~~~~~~~ +\widetilde{L}^4 \left(\widetilde{p}_r^2 \left(-\frac{27 \nu ^4}{64 \widetilde{r}^7}+\frac{81 \nu ^3}{4 \widetilde{r}^7}-\frac{603 \nu ^2}{64 \widetilde{r}^7}+\frac{15 \nu }{16 \widetilde{r}^7}\right)-\frac{11 \nu ^3}{8 \widetilde{r}^8}-\frac{449 \nu ^2}{32 \widetilde{r}^8}+\frac{29 \nu }{8 \widetilde{r}^8}\right)\nonumber\\
&~~~~~~~~~~ +\widetilde{L}^2 \bigg(\widetilde{p}_r^2 \left(\frac{325 \nu ^3}{32 \widetilde{r}^6}-\frac{7365 \nu ^2}{64 \widetilde{r}^6}+\frac{47 \nu }{4 \widetilde{r}^6}\right)+\widetilde{p}_r^4 \left(-\frac{69 \nu ^4}{64 \widetilde{r}^5}+\frac{2025 \nu ^3}{32 \widetilde{r}^5}-\frac{891 \nu ^2}{64 \widetilde{r}^5}\right)\nonumber\\
&~~~~~~~~~~~~~~~~~ +\frac{57 \nu ^3}{16 \widetilde{r}^7}+\frac{\left(12478-675 \pi ^2\right) \nu ^2}{1200 \widetilde{r}^7}+\frac{49 \nu }{4 \widetilde{r}^7}\bigg)\nonumber\\
&~~~~~~~~~~+\widetilde{p}_r^2 \left(\frac{145 \nu ^3}{8 \widetilde{r}^5}+\frac{\left(5400 \pi ^2-325199\right) \nu ^2}{2400 \widetilde{r}^5}+\frac{9 \nu }{2 \widetilde{r}^5}\right)+\widetilde{p}_r^4 \left(\frac{783 \nu ^3}{32 \widetilde{r}^4}+\frac{4887 \nu ^2}{64 \widetilde{r}^4}-\frac{31 \nu }{8 \widetilde{r}^4}\right)\nonumber\\
&~~~~~~~~~~+\widetilde{p}_r^6 \left(-\frac{187 \nu ^4}{64 \widetilde{r}^3}-\frac{253 \nu ^3}{16 \widetilde{r}^3}+\frac{93 \nu ^2}{16 \widetilde{r}^3}-\frac{5 \nu }{16 \widetilde{r}^3}\right)+\frac{\left(5296-6975 \pi ^2\right) \nu ^2}{1600 \widetilde{r}^6}+\frac{67 \nu }{2 \widetilde{r}^6}
\Bigg\}\nonumber\\
&+\left((\widetilde{\textbf{S}}_{(1)}\cdot\widetilde{\textbf{r}})(\widetilde{\textbf{S}}_{(2)}\cdot\widetilde{\textbf{r}})\right)\nonumber\\
&~~~~~~~~~~\Bigg\{
\widetilde{L}^6 \left(\frac{15 \nu ^4}{16 \widetilde{r}^{11}}-\frac{15 \nu ^3}{8 \widetilde{r}^{11}}+\frac{9 \nu ^2}{16 \widetilde{r}^{11}}\right)+\widetilde{L}^4 \left(\widetilde{p}_r^2 \left(\frac{225 \nu ^4}{64 \widetilde{r}^9}-\frac{779 \nu ^3}{32 \widetilde{r}^9}+\frac{69 \nu ^2}{8 \widetilde{r}^9}+\frac{5 \nu }{16 \widetilde{r}^9}\right)-\frac{63 \nu ^3}{8 \widetilde{r}^{10}}-\frac{5 \nu ^2}{\widetilde{r}^{10}}+\frac{17 \nu }{8 \widetilde{r}^{10}}\right) \nonumber\\
&~~~~~~~~~~+\widetilde{L}^2 \bigg(\widetilde{p}_r^2 \left(\frac{20 \nu ^3}{\widetilde{r}^8}+\frac{1005 \nu ^2}{8 \widetilde{r}^8}-\frac{3 \nu }{\widetilde{r}^8}\right)+\widetilde{p}_r^4 \left(\frac{177 \nu ^4}{32 \widetilde{r}^7}-\frac{1849 \nu ^3}{32 \widetilde{r}^7}+\frac{615 \nu ^2}{64 \widetilde{r}^7}+\frac{5 \nu }{8 \widetilde{r}^7}\right) \nonumber\\
&~~~~~~~~~~~~~~~~~ +\frac{79 \nu ^3}{16 \widetilde{r}^9}+\frac{\left(270 \pi ^2-17917\right) \nu ^2}{96 \widetilde{r}^9}+\frac{107 \nu }{4 \widetilde{r}^9}\bigg) \nonumber\\
&~~~~~~~~~~+\widetilde{p}_r^2 \left(\frac{327 \nu ^3}{8 \widetilde{r}^7}+\frac{\left(494789-5400 \pi ^2\right) \nu ^2}{800 \widetilde{r}^7}-\frac{83 \nu }{2 \widetilde{r}^7}\right)+\widetilde{p}_r^4 \left(-\frac{3919 \nu ^3}{32 \widetilde{r}^6}-\frac{3575 \nu ^2}{64 \widetilde{r}^6}+\frac{7 \nu }{8 \widetilde{r}^6}\right)\nonumber\\
&~~~~~~~~~~+\widetilde{p}_r^6 \left(\frac{699 \nu ^4}{64 \widetilde{r}^5}+\frac{125 \nu ^3}{16 \widetilde{r}^5}-\frac{69 \nu ^2}{16 \widetilde{r}^5}+\frac{5 \nu }{16 \widetilde{r}^5}\right)+\frac{\left(20925 \pi ^2-71168\right) \nu ^2}{1600 \widetilde{r}^8}-\frac{135 \nu }{2 \widetilde{r}^8}
\Bigg\}\nonumber\\
&+\left((\widetilde{\textbf{S}}_{(1)}\cdot\widetilde{\textbf{L}})(\widetilde{\textbf{S}}_{(2)}\cdot\widetilde{\textbf{L}})\right)\nonumber\\
&~~~~~~~~~~\Bigg\{
\widetilde{L}^4 \left(-\frac{15 \nu ^4}{16 \widetilde{r}^9}+\frac{59 \nu ^3}{16 \widetilde{r}^9}+\frac{39 \nu ^2}{8 \widetilde{r}^9}-\frac{25 \nu }{16 \widetilde{r}^9}\right)\nonumber\\
&~~~~~~~~~~+\widetilde{L}^2 \left(\widetilde{p}_r^2 \left(-\frac{27 \nu ^4}{8 \widetilde{r}^7}-\frac{25 \nu ^3}{2 \widetilde{r}^7}+\frac{489 \nu ^2}{32 \widetilde{r}^7}-\frac{25 \nu }{8 \widetilde{r}^7}\right)+\frac{387 \nu ^3}{16 \widetilde{r}^8}+\frac{1799 \nu ^2}{32 \widetilde{r}^8}-\frac{51 \nu }{4 \widetilde{r}^8}\right)\nonumber\\
&~~~~~~~~~~+\widetilde{p}_r^2 \left(-\frac{1285 \nu ^3}{16 \widetilde{r}^6}-\frac{3141 \nu ^2}{32 \widetilde{r}^6}-\frac{75 \nu }{4 \widetilde{r}^6}\right)+\widetilde{p}_r^4 \left(-\frac{69 \nu ^4}{16 \widetilde{r}^5}-\frac{677 \nu ^3}{8 \widetilde{r}^5}+\frac{933 \nu ^2}{32 \widetilde{r}^5}-\frac{25 \nu }{16 \widetilde{r}^5}\right)\nonumber\\
&~~~~~~~~~~-\frac{22 \nu ^3}{\widetilde{r}^7}+\frac{\left(43151-1350 \pi ^2\right) \nu ^2}{1200 \widetilde{r}^7}-\frac{221 \nu }{4 \widetilde{r}^7}
\Bigg\}\nonumber\\
&+\left((\widetilde{\textbf{S}}_{(1)}\cdot\widetilde{\textbf{r}})(\widetilde{\textbf{S}}_{(2)}\cdot\widetilde{\textbf{L}})\right) \nonumber\\
&~~~~~~~~~~\widetilde{p}_r \Bigg\{
\widetilde{L}^4 \left(-\frac{27 \nu ^4}{64 \widetilde{r}^9}-\frac{149 \nu ^3}{32 \widetilde{r}^9}+\frac{15 \nu ^2}{4 \widetilde{r}^9}-\frac{5 \nu }{8 \widetilde{r}^9}\right)\nonumber\\
&~~~~~~~~~~+\widetilde{L}^2 \left( \left(-\frac{\nu ^4}{2 \widetilde{r}^8}+\frac{1981 \nu ^3}{16 \widetilde{r}^8}-\frac{295 \nu ^2}{16 \widetilde{r}^8}-\frac{3 \nu }{4 \widetilde{r}^8}\right)+\widetilde{p}_r^2 \left(-\frac{45 \nu ^4}{32 \widetilde{r}^7}-\frac{1009 \nu ^3}{32 \widetilde{r}^7}+\frac{819 \nu ^2}{64 \widetilde{r}^7}-\frac{5 \nu }{4 \widetilde{r}^7}\right)\right)\nonumber\\
&~~~~~~~~~~-\frac{4593 \nu ^3}{32 \widetilde{r}^7}+\frac{\left(10800 \pi ^2-842053\right) \nu ^2}{2400 \widetilde{r}^7}+\frac{41 \nu }{2 \widetilde{r}^7} +\widetilde{p}_r^2 \left(-\frac{5 \nu ^4}{\widetilde{r}^6}-\frac{795 \nu ^3}{16 \widetilde{r}^6}+\frac{603 \nu ^2}{16 \widetilde{r}^6}-\frac{27 \nu }{4 \widetilde{r}^6}\right)\nonumber\\
&~~~~~~~~~~+\widetilde{p}_r^4 \left(\frac{27 \nu ^4}{64 \widetilde{r}^5}-\frac{55 \nu ^3}{16 \widetilde{r}^5}+\frac{51 \nu ^2}{16 \widetilde{r}^5}-\frac{5 \nu }{8 \widetilde{r}^5}\right) \nonumber\\
&~~~~~~~~~~+\frac{1}{q} \bigg(\widetilde{L}^4 \left(\frac{3 \nu ^4}{4 \widetilde{r}^9}+\frac{9 \nu ^3}{8 \widetilde{r}^9}+\frac{15 \nu ^2}{16 \widetilde{r}^9}\right) +\widetilde{L}^2 \left( \left(-\frac{\nu ^4}{2 \widetilde{r}^8}+\frac{3057 \nu ^3}{32 \widetilde{r}^8}+\frac{3 \nu ^2}{2 \widetilde{r}^8}\right)+\widetilde{p}_r^2\left(\frac{27 \nu ^4}{8 \widetilde{r}^7}-\frac{33 \nu ^3}{2 \widetilde{r}^7}+\frac{15 \nu ^2}{8 \widetilde{r}^7}\right)\right)\nonumber\\
&~~~~~~~~~~~~~~~~~+\frac{121 \nu ^2}{4 \widetilde{r}^7}-\frac{3227 \nu ^3}{32 \widetilde{r}^7}+\widetilde{p}_r^2 \left(-\frac{5 \nu ^4}{\widetilde{r}^6}-\frac{4269 \nu ^3}{32 \widetilde{r}^6}+\frac{27 \nu ^2}{2 \widetilde{r}^6}\right) +\widetilde{p}_r^4 \left(\frac{147 \nu ^4}{16 \widetilde{r}^5}-\frac{9 \nu ^3}{2 \widetilde{r}^5}+\frac{15 \nu ^2}{16 \widetilde{r}^5}\right)\bigg)
\Bigg\}\nonumber\\
&+\left((\widetilde{\textbf{S}}_{(1)}\cdot\widetilde{\textbf{L}})(\widetilde{\textbf{S}}_{(2)}\cdot\widetilde{\textbf{r}})\right)\nonumber\\
&~~~~~~~~~~ \widetilde{p}_r \Bigg\{
\widetilde{L}^4 \left(-\frac{123 \nu ^4}{64 \widetilde{r}^9}-\frac{197 \nu ^3}{32 \widetilde{r}^9}+\frac{3 \nu ^2}{\widetilde{r}^9}+\frac{5 \nu }{16 \widetilde{r}^9}\right)\nonumber\\
&~~~~~~~~~~+\widetilde{L}^2 \left(\left(\frac{\nu ^4}{2 \widetilde{r}^8}-\frac{271 \nu ^3}{4 \widetilde{r}^8}+\frac{2371 \nu ^2}{32 \widetilde{r}^8}+\frac{3 \nu }{4 \widetilde{r}^8}\right)+\widetilde{p}_r^2 \left(-\frac{261 \nu ^4}{32 \widetilde{r}^7}+\frac{155 \nu ^3}{32 \widetilde{r}^7}-\frac{477 \nu ^2}{64 \widetilde{r}^7}+\frac{5 \nu }{8 \widetilde{r}^7}\right)\right)\nonumber\\
&~~~~~~~~~~+\frac{1861 \nu ^3}{32 \widetilde{r}^7}+\frac{\left(5400 \pi ^2-614639\right) \nu ^2}{1200 \widetilde{r}^7}+\frac{203 \nu }{4 \widetilde{r}^7}+\widetilde{p}_r^2 \left(\frac{5 \nu ^4}{\widetilde{r}^6}+\frac{1697 \nu ^3}{8 \widetilde{r}^6}-\frac{3927 \nu ^2}{32 \widetilde{r}^6}+\frac{27 \nu }{4 \widetilde{r}^6}\right)\nonumber\\
&~~~~~~~~~~+\widetilde{p}_r^4 \left(-\frac{1149 \nu ^4}{64 \widetilde{r}^5}+\frac{59 \nu ^3}{4 \widetilde{r}^5}-\frac{51 \nu ^2}{16 \widetilde{r}^5}+\frac{5 \nu }{16 \widetilde{r}^5}\right)\nonumber\\
&~~~~~~~~~~+\frac{1}{q} \bigg(\widetilde{L}^4 \left(-\frac{3 \nu ^4}{4 \widetilde{r}^9}-\frac{9 \nu ^3}{8 \widetilde{r}^9}-\frac{15 \nu ^2}{16 \widetilde{r}^9}\right) +\widetilde{L}^2 \left( \left(\frac{\nu ^4}{2 \widetilde{r}^8}-\frac{3057 \nu ^3}{32 \widetilde{r}^8}-\frac{3 \nu ^2}{2 \widetilde{r}^8}\right)+\widetilde{p}_r^2 \left(-\frac{27 \nu ^4}{8 \widetilde{r}^7}+\frac{33 \nu ^3}{2 \widetilde{r}^7}-\frac{15 \nu ^2}{8 \widetilde{r}^7}\right)\right)
\nonumber\\
&~~~~~~~~~~~~~~~~~~+ \left(\frac{3227 \nu ^3}{32 \widetilde{r}^7}-\frac{121 \nu ^2}{4 \widetilde{r}^7}\right)+\widetilde{p}_r^2 \left(\frac{5 \nu ^4}{\widetilde{r}^6}+\frac{4269 \nu ^3}{32 \widetilde{r}^6}-\frac{27 \nu ^2}{2 \widetilde{r}^6}\right) +\widetilde{p}_r^4 \left(-\frac{147 \nu ^4}{16 \widetilde{r}^5}+\frac{9 \nu ^3}{2 \widetilde{r}^5}-\frac{15 \nu ^2}{16 \widetilde{r}^5}\right)\bigg)
\Bigg\}
\end{align}
\begin{align}
\widetilde{\mathcal{H}}^{\rm S^2}_{\rm N^3LO}&= 
\left(\widetilde{\textbf{S}}_{(1)}\cdot\widetilde{\textbf{S}}_{(1)}\right)\nonumber\\
&~~~~ \Bigg\{
\widetilde{L}^6 \left(\frac{67 \nu ^4}{32 \widetilde{r}^9}-\frac{499 \nu ^3}{128 \widetilde{r}^9}+\frac{47 \nu ^2}{64 \widetilde{r}^9}\right) +\widetilde{L}^4 \left(\widetilde{p}_r^2 \left(\frac{1113 \nu ^4}{128 \widetilde{r}^7}-\frac{21 \nu ^3}{16 \widetilde{r}^7}+\frac{99 \nu ^2}{128 \widetilde{r}^7}\right)-\frac{169 \nu ^4}{128 \widetilde{r}^8}-\frac{5027 \nu ^3}{128 \widetilde{r}^8}+\frac{947 \nu ^2}{128 \widetilde{r}^8}\right)\nonumber\\
&~~~~ +\widetilde{L}^2 \bigg(\widetilde{p}_r^2 \left(-\frac{247 \nu ^4}{64 \widetilde{r}^6}+\frac{3179 \nu ^3}{64 \widetilde{r}^6}-\frac{15 \nu ^2}{64 \widetilde{r}^6}\right)+\widetilde{p}_r^4 \left(\frac{801 \nu ^4}{64 \widetilde{r}^5}-\frac{669 \nu ^3}{128 \widetilde{r}^5}-\frac{21 \nu ^2}{32 \widetilde{r}^5}\right)+\frac{\left(21311+225 \pi ^2\right) \nu ^3}{600 \widetilde{r}^7}+\frac{48541 \nu ^2}{1960 \widetilde{r}^7}\bigg)\nonumber\\
&~~~~+\widetilde{p}_r^2 \left(-\frac{\left(121757+25200 \pi ^2\right) \nu ^3}{16800 \widetilde{r}^5}-\frac{25553 \nu ^2}{3920 \widetilde{r}^5}\right)+\widetilde{p}_r^4 \left(\frac{587 \nu ^4}{128 \widetilde{r}^4}+\frac{5187 \nu ^3}{128 \widetilde{r}^4}-\frac{785 \nu ^2}{128 \widetilde{r}^4}\right)\nonumber\\
&~~~~+\widetilde{p}_r^6 \left(-\frac{293 \nu ^4}{128 \widetilde{r}^3}+\frac{235 \nu ^3}{64 \widetilde{r}^3}-\frac{89 \nu ^2}{128 \widetilde{r}^3}\right)+\frac{23913 \nu ^2}{3920 \widetilde{r}^6}-\frac{\left(21619+3150 \pi ^2\right) \nu ^3}{8400 \widetilde{r}^6}\nonumber\\
&~~~~+\frac{1}{q}\Bigg( \widetilde{L}^6 \left(\frac{253 \nu ^4}{128 \widetilde{r}^9}-\frac{75 \nu ^3}{8 \widetilde{r}^9}+\frac{861 \nu ^2}{128 \widetilde{r}^9}-\frac{47 \nu }{64 \widetilde{r}^9}\right)\nonumber\\
&~~~~~~~~~~+\widetilde{L}^4 \left(\widetilde{p}_r^2 \left(\frac{261 \nu ^4}{32 \widetilde{r}^7}-\frac{831 \nu ^3}{64 \widetilde{r}^7}+\frac{165 \nu ^2}{64 \widetilde{r}^7}-\frac{99 \nu }{128 \widetilde{r}^7}\right)-\frac{169 \nu ^4}{128 \widetilde{r}^8}-\frac{2123 \nu ^3}{64 \widetilde{r}^8}+\frac{1531 \nu ^2}{128 \widetilde{r}^8}-\frac{899 \nu }{128 \widetilde{r}^8}\right)\nonumber\\
&~~~~~~~~~~+\widetilde{L}^2 \bigg(\widetilde{p}_r^2 \left(-\frac{247 \nu ^4}{64 \widetilde{r}^6}+\frac{4319 \nu ^3}{64 \widetilde{r}^6}+\frac{7103 \nu ^2}{64 \widetilde{r}^6}+\frac{7 \nu }{64 \widetilde{r}^6}\right)+\widetilde{p}_r^4 \left(\frac{1479 \nu ^4}{128 \widetilde{r}^5}-\frac{9 \nu ^3}{8 \widetilde{r}^5}-\frac{1083 \nu ^2}{128 \widetilde{r}^5}+\frac{21 \nu }{32 \widetilde{r}^5}\right)\nonumber\\
&~~~~~~~~~~~~~~~~+\frac{\left(149027+1575 \pi ^2\right) \nu ^3}{4200 \widetilde{r}^7}+\frac{\left(13568+1341 \pi ^2\right) \nu ^2}{6144 \widetilde{r}^7}-\frac{221 \nu }{8 \widetilde{r}^7}\bigg)\nonumber\\
&~~~~~~~~~~+\widetilde{p}_r^2 \left(-\frac{\left(38207+25200 \pi ^2\right) \nu ^3}{16800 \widetilde{r}^5}+\frac{\left(3952-1341 \pi ^2\right) \nu ^2}{1536 \widetilde{r}^5}+\frac{185 \nu }{16 \widetilde{r}^5}\right)\nonumber\\
&~~~~~~~~~~+\widetilde{p}_r^4 \left(\frac{587 \nu ^4}{128 \widetilde{r}^4}+\frac{3949 \nu ^3}{64 \widetilde{r}^4}-\frac{2101 \nu ^2}{128 \widetilde{r}^4}+\frac{721 \nu }{128 \widetilde{r}^4}\right)+\widetilde{p}_r^6 \left(-\frac{181 \nu ^4}{64 \widetilde{r}^3}+\frac{369 \nu ^3}{64 \widetilde{r}^3}-\frac{69 \nu ^2}{16 \widetilde{r}^3}+\frac{89 \nu }{128 \widetilde{r}^3}\right)\nonumber\\
&~~~~~~~~~~-\frac{\left(21619+3150 \pi ^2\right) \nu ^3}{8400 \widetilde{r}^6}+\frac{\left(587-9 \pi ^2\right) \nu ^2}{48 \widetilde{r}^6}+\frac{155 \nu }{16 \widetilde{r}^6}
\Bigg)
\Bigg\}\nonumber\\
&+\left((\widetilde{\textbf{S}}_{(1)}\cdot\widetilde{\textbf{r}})^2\right)\nonumber\\
&~~~~\Bigg\{
\widetilde{L}^4 \left(\widetilde{p}_r^2 \left(-\frac{133 \nu ^4}{128 \widetilde{r}^9}-\frac{265 \nu ^3}{64 \widetilde{r}^9}+\frac{41 \nu ^2}{128 \widetilde{r}^9}\right)-\frac{113 \nu ^4}{128 \widetilde{r}^{10}}-\frac{795 \nu ^3}{128 \widetilde{r}^{10}}+\frac{159 \nu ^2}{128 \widetilde{r}^{10}}\right) \nonumber\\
&~~~~+\widetilde{L}^2 \bigg(\widetilde{p}_r^2 \left(-\frac{93 \nu ^4}{32 \widetilde{r}^8}-\frac{1005 \nu ^3}{32 \widetilde{r}^8}+\frac{61 \nu ^2}{16 \widetilde{r}^8}\right)+\widetilde{p}_r^4 \left(-\frac{145 \nu ^4}{64 \widetilde{r}^7}+\frac{35 \nu ^3}{32 \widetilde{r}^7}+\frac{65 \nu ^2}{64 \widetilde{r}^7}\right)+\frac{\left(1832-315 \pi ^2\right) \nu ^3}{168 \widetilde{r}^9}+\frac{22779 \nu ^2}{784 \widetilde{r}^9}\bigg) \nonumber\\
&~~~~+\widetilde{p}_r^2\left(\frac{\left(4927+25200 \pi ^2\right) \nu ^3}{5600 \widetilde{r}^7}-\frac{118151 \nu ^2}{3920 \widetilde{r}^7}\right)+\widetilde{p}_r^4 \left(-\frac{1147 \nu ^4}{128 \widetilde{r}^6}-\frac{2483 \nu ^3}{128 \widetilde{r}^6}+\frac{481 \nu ^2}{128 \widetilde{r}^6}\right)\nonumber\\
&~~~~+\widetilde{p}_r^6 \left(\frac{293 \nu ^4}{128 \widetilde{r}^5}-\frac{235 \nu ^3}{64 \widetilde{r}^5}+\frac{89 \nu ^2}{128 \widetilde{r}^5}\right)+\frac{\left(3150 \pi ^2-11071\right) \nu ^3}{2800 \widetilde{r}^8}-\frac{71809 \nu ^2}{3920 \widetilde{r}^8}\nonumber\\
&~~~~+\frac{1}{q}\Bigg(
\widetilde{L}^4 \left(\widetilde{p}_r^2 \left(-\frac{59 \nu ^4}{64 \widetilde{r}^9}-\frac{327 \nu ^3}{64 \widetilde{r}^9}+\frac{15 \nu ^2}{8 \widetilde{r}^9}-\frac{41 \nu }{128 \widetilde{r}^9}\right)-\frac{113 \nu ^4}{128 \widetilde{r}^{10}}-\frac{101 \nu ^3}{16 \widetilde{r}^{10}}+\frac{267 \nu ^2}{128 \widetilde{r}^{10}}-\frac{31 \nu }{128 \widetilde{r}^{10}}\right)\nonumber\\
&~~~~~~~~~~+\widetilde{L}^2 \bigg(\widetilde{p}_r^2 \left(-\frac{93 \nu ^4}{32 \widetilde{r}^8}-\frac{1633 \nu ^3}{64 \widetilde{r}^8}-\frac{573 \nu ^2}{32 \widetilde{r}^8}-\frac{75 \nu }{16 \widetilde{r}^8}\right)+\widetilde{p}_r^4 \left(-\frac{59 \nu ^4}{32 \widetilde{r}^7}-\frac{183 \nu ^3}{32 \widetilde{r}^7}+\frac{159 \nu ^2}{16 \widetilde{r}^7}-\frac{65 \nu }{64 \widetilde{r}^7}\right)\nonumber\\
&~~~~~~~~~~~~~~~~+\frac{\left(7841-1260 \pi ^2\right) \nu ^3}{672 \widetilde{r}^9}+\frac{\left(235904-6705 \pi ^2\right) \nu ^2}{6144 \widetilde{r}^9}-\frac{35 \nu }{16 \widetilde{r}^9}\bigg)\nonumber\\
&~~~~~~~~~~+\widetilde{p}_r^2 \left(\frac{9 \left(7053+2800 \pi ^2\right) \nu ^3}{5600 \widetilde{r}^7}+\frac{3 \left(447 \pi ^2-2032\right) \nu ^2}{512 \widetilde{r}^7}-\frac{185 \nu }{16 \widetilde{r}^7}\right)\nonumber\\
&~~~~~~~~~~+\widetilde{p}_r^4 \left(-\frac{1147 \nu ^4}{128 \widetilde{r}^6}-\frac{3221 \nu ^3}{64 \widetilde{r}^6}+\frac{2101 \nu ^2}{128 \widetilde{r}^6}-\frac{721 \nu }{128 \widetilde{r}^6}\right)+\widetilde{p}_r^6 \left(\frac{181 \nu ^4}{64 \widetilde{r}^5}-\frac{369 \nu ^3}{64 \widetilde{r}^5}+\frac{69 \nu ^2}{16 \widetilde{r}^5}-\frac{89 \nu }{128 \widetilde{r}^5}\right)\nonumber\\
&~~~~~~~~~~+\frac{\left(3150 \pi ^2-11071\right) \nu ^3}{2800 \widetilde{r}^8}+\frac{3 \left(3 \pi ^2-179\right) \nu ^2}{16 \widetilde{r}^8}-\frac{155 \nu }{16 \widetilde{r}^8}
\Bigg)
\Bigg\}\nonumber\\
&+\left((\widetilde{\textbf{S}}_{(1)}\cdot\widetilde{\textbf{L}})^2\right)\nonumber\\
&~~~~\Bigg\{
\widetilde{L}^4 \left(-\frac{661 \nu ^4}{128 \widetilde{r}^9}+\frac{1087 \nu ^3}{128 \widetilde{r}^9}-\frac{193 \nu ^2}{128 \widetilde{r}^9}\right)+\widetilde{L}^2 \left(\widetilde{p}_r^2 \left(-\frac{1423 \nu ^4}{64 \widetilde{r}^7}+\frac{697 \nu ^3}{64 \widetilde{r}^7}-\frac{169 \nu ^2}{64 \widetilde{r}^7}\right)+\frac{297 \nu ^4}{64 \widetilde{r}^8}+\frac{5345 \nu ^3}{64 \widetilde{r}^8}-\frac{1217 \nu ^2}{64 \widetilde{r}^8}\right)\nonumber\\
&~~~~+\widetilde{p}_r^2 \left(\frac{1005 \nu ^4}{64 \widetilde{r}^6}+\frac{3277 \nu ^3}{64 \widetilde{r}^6}-\frac{1141 \nu ^2}{64 \widetilde{r}^6}\right)+\widetilde{p}_r^4 \left(-\frac{4675 \nu ^4}{128 \widetilde{r}^5}+\frac{2317 \nu ^3}{128 \widetilde{r}^5}-\frac{145 \nu ^2}{128 \widetilde{r}^5}\right)\nonumber\\
&~~~~+\frac{\left(12600 \pi ^2-1208189\right) \nu ^3}{16800 \widetilde{r}^7}-\frac{114069 \nu ^2}{980 \widetilde{r}^7}\nonumber\\
&~~~~+\frac{1}{q}\Bigg(
\widetilde{L}^4 \left(-\frac{601 \nu ^4}{128 \widetilde{r}^9}+\frac{1431 \nu ^3}{64 \widetilde{r}^9}-\frac{1695 \nu ^2}{128 \widetilde{r}^9}+\frac{193 \nu }{128 \widetilde{r}^9}\right)\nonumber\\
&~~~~~~~~~~+\widetilde{L}^2 \left(\widetilde{p}_r^2 \left(-\frac{1291 \nu ^4}{64 \widetilde{r}^7}+\frac{375 \nu ^3}{8 \widetilde{r}^7}-\frac{1119 \nu ^2}{64 \widetilde{r}^7}+\frac{169 \nu }{64 \widetilde{r}^7}\right)+\frac{297 \nu ^4}{64 \widetilde{r}^8}+\frac{2135 \nu ^3}{32 \widetilde{r}^8}-\frac{3005 \nu ^2}{64 \widetilde{r}^8}+\frac{1089 \nu }{64 \widetilde{r}^8}\right)\nonumber\\
&~~~~~~~~~~+\widetilde{p}_r^2 \left(\frac{1005 \nu ^4}{64 \widetilde{r}^6}-\frac{331 \nu ^3}{32 \widetilde{r}^6}-\frac{21555 \nu ^2}{64 \widetilde{r}^6}+\frac{1013 \nu }{64 \widetilde{r}^6}\right)+\widetilde{p}_r^4 \left(-\frac{4231 \nu ^4}{128 \widetilde{r}^5}+\frac{1449 \nu ^3}{64 \widetilde{r}^5}-\frac{1023 \nu ^2}{128 \widetilde{r}^5}+\frac{145 \nu }{128 \widetilde{r}^5}\right)\nonumber\\
&~~~~~~~~~~+\frac{\left(12600 \pi ^2-1175639\right) \nu ^3}{16800 \widetilde{r}^7}+\frac{\left(1341 \pi ^2-317152\right) \nu ^2}{3072 \widetilde{r}^7}+\frac{353 \nu }{4 \widetilde{r}^7}
\Bigg)
\Bigg\}\nonumber\\
&+\left((\widetilde{\textbf{S}}_{(1)}\cdot\widetilde{\textbf{r}})(\widetilde{\textbf{S}}_{(1)}\cdot\widetilde{\textbf{L}})\right) \nonumber\\
&~~~~\widetilde{p}_r \Bigg\{
\widetilde{L}^4 \left(-\frac{401 \nu ^4}{128 \widetilde{r}^9}-\frac{31 \nu ^3}{128 \widetilde{r}^9}-\frac{53 \nu ^2}{128 \widetilde{r}^9}\right)+\widetilde{L}^2 \left(\left(\frac{65 \nu ^4}{16 \widetilde{r}^8}+\frac{55 \nu ^3}{4 \widetilde{r}^8}-\frac{25 \nu ^2}{16 \widetilde{r}^8}\right)+\widetilde{p}_r^2 \left(-\frac{635 \nu ^4}{64 \widetilde{r}^7}+\frac{419 \nu ^3}{64 \widetilde{r}^7}-\frac{5 \nu ^2}{64 \widetilde{r}^7}\right)\right)\nonumber\\
&~~~~~~~+\widetilde{p}_r^2 \left(\frac{185 \nu ^4}{16 \widetilde{r}^6}-\frac{1219 \nu ^3}{16 \widetilde{r}^6}+\frac{37 \nu ^2}{16 \widetilde{r}^6}\right)+\widetilde{p}_r^4 \left(-\frac{1019 \nu ^4}{128 \widetilde{r}^5}+\frac{59 \nu ^3}{128 \widetilde{r}^5}+\frac{43 \nu ^2}{128 \widetilde{r}^5}\right) \nonumber\\
&~~~~~~~+\frac{109993 \nu ^2}{3920 \widetilde{r}^7}-\frac{\left(155101+12600 \pi ^2\right) \nu ^3}{2100 \widetilde{r}^7} \nonumber\\
&~~~~~~~+\frac{1}{q} \bigg(\widetilde{L}^4 \left(-\frac{371 \nu ^4}{128 \widetilde{r}^9}+\frac{273 \nu ^3}{64 \widetilde{r}^9}-\frac{621 \nu ^2}{128 \widetilde{r}^9}+\frac{53 \nu }{128 \widetilde{r}^9}\right)\nonumber\\
&~~~~~~~~~~~~+\widetilde{L}^2 \left(\left(\frac{65 \nu ^4}{16 \widetilde{r}^8}+\frac{13 \nu ^3}{2 \widetilde{r}^8}-\frac{549 \nu ^2}{16 \widetilde{r}^8}+\frac{49 \nu }{16 \widetilde{r}^8}\right)+\widetilde{p}_r^2 \left(-\frac{581 \nu ^4}{64 \widetilde{r}^7}+\frac{99 \nu ^3}{8 \widetilde{r}^7}+\frac{351 \nu ^2}{64 \widetilde{r}^7}+\frac{5 \nu }{64 \widetilde{r}^7}\right)\right)\nonumber\\
&~~~~~~~~~~~~+\widetilde{p}_r^2 \left(\frac{185 \nu ^4}{16 \widetilde{r}^6}-\frac{1093 \nu ^3}{16 \widetilde{r}^6}-\frac{1383 \nu ^2}{16 \widetilde{r}^6}-\frac{13 \nu }{16 \widetilde{r}^6}\right)+\widetilde{p}_r^4 \left(-\frac{881 \nu ^4}{128 \widetilde{r}^5}+\frac{69 \nu ^3}{64 \widetilde{r}^5}+\frac{363 \nu ^2}{128 \widetilde{r}^5}-\frac{43 \nu }{128 \widetilde{r}^5}\right)\nonumber\\
&~~~~~~~~~~~~-\frac{\left(729829+50400 \pi ^2\right) \nu ^3}{8400 \widetilde{r}^7}+\frac{\left(9544-1341 \pi ^2\right) \nu ^2}{384 \widetilde{r}^7}+\frac{315 \nu }{16 \widetilde{r}^7}
\bigg)
\Bigg\}\nonumber\\
&+(1\leftrightarrow 2)
\end{align}
\begin{align}
\widetilde{\mathcal{H}}^{\rm ES^2}_{\rm N^3LO}&= 
\left(C^{(0)}_{\rm ES^2}\right)_{(1)} \left(\widetilde{\textbf{S}}_{(1)}\cdot\widetilde{\textbf{S}}_{(1)}\right)\nonumber\\
&~~~~ \Bigg\{
\widetilde{L}^6 \left(-\frac{15 \nu ^4}{16 \widetilde{r}^9}+\frac{\nu ^3}{\widetilde{r}^9}-\frac{3 \nu ^2}{16 \widetilde{r}^9}\right) +\widetilde{L}^4 \left(\widetilde{p}_r^2 \left(\frac{9 \nu ^3}{4 \widetilde{r}^7}-\frac{27 \nu ^4}{8 \widetilde{r}^7}\right)+\frac{9 \nu ^4}{16 \widetilde{r}^8}-\frac{69 \nu ^3}{16 \widetilde{r}^8}-\frac{221 \nu ^2}{16 \widetilde{r}^8}\right)\nonumber\\
&~~~~ +\widetilde{L}^2 \bigg(\widetilde{p}_r^2 \left(\frac{9 \nu ^4}{8 \widetilde{r}^6}+\frac{259 \nu ^3}{4 \widetilde{r}^6}+\frac{76 \nu ^2}{\widetilde{r}^6}\right)+\widetilde{p}_r^4 \left(-\frac{3 \nu ^4}{\widetilde{r}^5}-\frac{9 \nu ^3}{4 \widetilde{r}^5}+\frac{9 \nu ^2}{16 \widetilde{r}^5}\right)+\frac{447 \nu ^3}{56 \widetilde{r}^7}+\frac{125787 \nu ^2}{4900 \widetilde{r}^7}\bigg)\nonumber\\
&~~~~+\widetilde{p}_r^2 \left(-\frac{123 \nu ^3}{14 \widetilde{r}^5}-\frac{192173 \nu ^2}{4900 \widetilde{r}^5}\right)+\widetilde{p}_r^4 \left(-\frac{63 \nu ^4}{16 \widetilde{r}^4}-\frac{755 \nu ^3}{16 \widetilde{r}^4}-\frac{99 \nu ^2}{16 \widetilde{r}^4}\right)\nonumber\\
&~~~~+\widetilde{p}_r^6 \left(\frac{6 \nu ^4}{\widetilde{r}^3}-\frac{7 \nu ^3}{2 \widetilde{r}^3}+\frac{3 \nu ^2}{8 \widetilde{r}^3}\right)-\frac{69 \nu ^3}{28 \widetilde{r}^6}-\frac{68387 \nu ^2}{4900 \widetilde{r}^6}\nonumber\\
&~~~~+\frac{1}{q}\Bigg( \widetilde{L}^6 \left(-\frac{35 \nu ^4}{32 \widetilde{r}^9}-\frac{53 \nu ^3}{32 \widetilde{r}^9}+\frac{31 \nu ^2}{16 \widetilde{r}^9}-\frac{13 \nu }{32 \widetilde{r}^9}\right)\nonumber\\
&~~~~~~~~~~+\widetilde{L}^4 \left(\widetilde{p}_r^2 \left(-\frac{33 \nu ^4}{8 \widetilde{r}^7}-\frac{207 \nu ^3}{32 \widetilde{r}^7}+\frac{99 \nu ^2}{16 \widetilde{r}^7}-\frac{21 \nu }{32 \widetilde{r}^7}\right)+\frac{9 \nu ^4}{16 \widetilde{r}^8}+\frac{27 \nu ^3}{8 \widetilde{r}^8}+\frac{53 \nu ^2}{16 \widetilde{r}^8}-\frac{9 \nu }{2 \widetilde{r}^8}\right)\nonumber\\
&~~~~~~~~~~+\widetilde{L}^2 \bigg(\widetilde{p}_r^2 \left(\frac{9 \nu ^4}{8 \widetilde{r}^6}+\frac{365 \nu ^3}{4 \widetilde{r}^6}+\frac{1701 \nu ^2}{16 \widetilde{r}^6}-\frac{3 \nu }{\widetilde{r}^6}\right)+\widetilde{p}_r^4 \left(-\frac{9 \nu ^4}{2 \widetilde{r}^5}-\frac{75 \nu ^3}{8 \widetilde{r}^5}+\frac{45 \nu ^2}{16 \widetilde{r}^5}-\frac{3 \nu }{32 \widetilde{r}^5}\right)\nonumber\\
&~~~~~~~~~~~~~~~~+\frac{499 \nu ^3}{112 \widetilde{r}^7}+\frac{\left(177536-2547 \pi ^2\right) \nu ^2}{6144 \widetilde{r}^7}-\frac{205 \nu }{8 \widetilde{r}^7}\bigg)\nonumber\\
&~~~~~~~~~~+\widetilde{p}_r^2 \left(-\frac{457 \nu ^3}{28 \widetilde{r}^5}+\frac{\left(2547 \pi ^2-273728\right) \nu ^2}{1536 \widetilde{r}^5}+\frac{39 \nu }{8 \widetilde{r}^5}\right)\nonumber\\
&~~~~~~~~~~+\widetilde{p}_r^4 \left(-\frac{63 \nu ^4}{16 \widetilde{r}^4}-\frac{257 \nu ^3}{8 \widetilde{r}^4}-\frac{81 \nu ^2}{8 \widetilde{r}^4}+\frac{3 \nu }{2 \widetilde{r}^4}\right)+\widetilde{p}_r^6 \left(\frac{4 \nu ^4}{\widetilde{r}^3}+\frac{2 \nu ^3}{\widetilde{r}^3}-\frac{23 \nu ^2}{16 \widetilde{r}^3}+\frac{5 \nu }{32 \widetilde{r}^3}\right)\nonumber\\
&~~~~~~~~~~-\frac{69 \nu ^3}{28 \widetilde{r}^6}-\frac{\left(2888+387 \pi ^2\right) \nu ^2}{192 \widetilde{r}^6}+\frac{21 \nu }{2 \widetilde{r}^6}
\Bigg)
\Bigg\}\nonumber\\
&+\left(C^{(0)}_{\rm ES^2}\right)_{(1)}\left((\widetilde{\textbf{S}}_{(1)}\cdot\widetilde{\textbf{r}})^2\right)\nonumber\\
&~~~~\Bigg\{
\widetilde{L}^4 \left(\widetilde{p}_r^2 \left(-\frac{9 \nu ^4}{16 \widetilde{r}^9}-\frac{11 \nu ^3}{16 \widetilde{r}^9}+\frac{3 \nu ^2}{4 \widetilde{r}^9}\right)+\frac{15 \nu ^4}{16 \widetilde{r}^{10}}+\frac{311 \nu ^3}{16 \widetilde{r}^{10}}+\frac{33 \nu ^2}{16 \widetilde{r}^{10}}\right) \nonumber\\
&~~~~+\widetilde{L}^2 \bigg(\widetilde{p}_r^2 \left(\frac{35 \nu ^4}{8 \widetilde{r}^8}+\frac{107 \nu ^3}{4 \widetilde{r}^8}-\frac{37 \nu ^2}{2 \widetilde{r}^8}\right)+\widetilde{p}_r^4 \left(-\frac{45 \nu ^4}{16 \widetilde{r}^7}+\frac{17 \nu ^3}{4 \widetilde{r}^7}-\frac{21 \nu ^2}{8 \widetilde{r}^7}\right)-\frac{1063 \nu ^3}{56 \widetilde{r}^9}-\frac{3069 \nu ^2}{49 \widetilde{r}^9}\bigg) \nonumber\\
&~~~~+\widetilde{p}_r^2 \left(-\frac{1535 \nu ^3}{14 \widetilde{r}^7}-\frac{19671 \nu ^2}{4900 \widetilde{r}^7}\right)+\widetilde{p}_r^4 \left(\frac{203 \nu ^4}{16 \widetilde{r}^6}+\frac{263 \nu ^3}{16 \widetilde{r}^6}-\frac{457 \nu ^2}{16 \widetilde{r}^6}\right)\nonumber\\
&~~~~+\widetilde{p}_r^6 \left(-\frac{6 \nu ^4}{\widetilde{r}^5}+\frac{23 \nu ^3}{2 \widetilde{r}^5}-\frac{27 \nu ^2}{8 \widetilde{r}^5}\right)+\frac{111851 \nu ^2}{4900 \widetilde{r}^8}-\frac{101 \nu ^3}{28 \widetilde{r}^8}\nonumber\\
&~~~~+\frac{1}{q}\Bigg(
\widetilde{L}^6 \left(\frac{15 \nu ^4}{32 \widetilde{r}^{11}}+\frac{105 \nu ^3}{32 \widetilde{r}^{11}}-\frac{51 \nu ^2}{16 \widetilde{r}^{11}}+\frac{21 \nu }{32 \widetilde{r}^{11}}\right)\nonumber\\
&~~~~~~~~~~+\widetilde{L}^4 \left(\widetilde{p}_r^2 \left(\frac{21 \nu ^4}{16 \widetilde{r}^9}+\frac{319 \nu ^3}{32 \widetilde{r}^9}-\frac{73 \nu ^2}{8 \widetilde{r}^9}+\frac{39 \nu }{32 \widetilde{r}^9}\right)+\frac{15 \nu ^4}{16 \widetilde{r}^{10}}+\frac{10 \nu ^3}{\widetilde{r}^{10}}-\frac{129 \nu ^2}{8 \widetilde{r}^{10}}+\frac{15 \nu }{2 \widetilde{r}^{10}}\right)\nonumber\\
&~~~~~~~~~~+\widetilde{L}^2 \bigg(\widetilde{p}_r^2 \left(\frac{35 \nu ^4}{8 \widetilde{r}^8}-\frac{163 \nu ^3}{16 \widetilde{r}^8}-\frac{1931 \nu ^2}{16 \widetilde{r}^8}+\frac{6 \nu }{\widetilde{r}^8}\right)+\widetilde{p}_r^4 \left(\frac{3 \nu ^4}{8 \widetilde{r}^7}+\frac{181 \nu ^3}{16 \widetilde{r}^7}-\frac{73 \nu ^2}{16 \widetilde{r}^7}+\frac{3 \nu }{32 \widetilde{r}^7}\right)\nonumber\\
&~~~~~~~~~~~~~~~~+-\frac{1725 \nu ^3}{112 \widetilde{r}^9}+\frac{\left(12735 \pi ^2-929536\right) \nu ^2}{6144 \widetilde{r}^9}+\frac{369 \nu }{8 \widetilde{r}^9}\bigg)\nonumber\\
&~~~~~~~~~~+\widetilde{p}_r^2 \left(-\frac{2493 \nu ^3}{28 \widetilde{r}^7}-\frac{3 \left(849 \pi ^2-59072\right) \nu ^2}{512 \widetilde{r}^7}-\frac{117 \nu }{8 \widetilde{r}^7}\right)\nonumber\\
&~~~~~~~~~~+\widetilde{p}_r^4 \left(\frac{203 \nu ^4}{16 \widetilde{r}^6}-\frac{44 \nu ^3}{\widetilde{r}^6}+\frac{7 \nu ^2}{8 \widetilde{r}^6}-\frac{9 \nu }{2 \widetilde{r}^6}\right)+\widetilde{p}_r^6 \left(-\frac{\nu ^3}{\widetilde{r}^5}+\frac{37 \nu ^2}{16 \widetilde{r}^5}-\frac{15 \nu }{32 \widetilde{r}^5}\right)\nonumber\\
&~~~~~~~~~~-\frac{101 \nu ^3}{28 \widetilde{r}^8}+\frac{\left(296+387 \pi ^2\right) \nu ^2}{64 \widetilde{r}^8}-\frac{63 \nu }{2 \widetilde{r}^8}
\Bigg)
\Bigg\}\nonumber\\
&+\left(C^{(0)}_{\rm ES^2}\right)_{(1)}\left((\widetilde{\textbf{S}}_{(1)}\cdot\widetilde{\textbf{L}})^2\right)\nonumber\\
&~~~~\Bigg\{
\widetilde{L}^4 \left(\frac{15 \nu ^4}{16 \widetilde{r}^9}-\frac{7 \nu ^3}{2 \widetilde{r}^9}+\frac{27 \nu ^2}{16 \widetilde{r}^9}\right)+\widetilde{L}^2 \left(\widetilde{p}_r^2 \left(\frac{63 \nu ^4}{16 \widetilde{r}^7}-\frac{97 \nu ^3}{16 \widetilde{r}^7}-\frac{3 \nu ^2}{4 \widetilde{r}^7}\right)-\frac{3 \nu ^4}{4 \widetilde{r}^8}+\frac{47 \nu ^3}{8 \widetilde{r}^8}+\frac{75 \nu ^2}{4 \widetilde{r}^8}\right)\nonumber\\
&~~~~+\widetilde{p}_r^2 \left(-\frac{2 \nu ^4}{\widetilde{r}^6}-\frac{351 \nu ^3}{4 \widetilde{r}^6}-\frac{245 \nu ^2}{4 \widetilde{r}^6}\right)+\widetilde{p}_r^4 \left(\frac{93 \nu ^4}{16 \widetilde{r}^5}+\frac{4 \nu ^3}{\widetilde{r}^5}-\frac{39 \nu ^2}{16 \widetilde{r}^5}\right)+\frac{35099 \nu ^2}{4900 \widetilde{r}^7}-\frac{73 \nu ^3}{14 \widetilde{r}^7}\nonumber\\
&~~~~+\frac{1}{q}\Bigg(
\widetilde{L}^4 \left(\frac{15 \nu ^4}{16 \widetilde{r}^9}-\frac{\nu ^3}{16 \widetilde{r}^9}-\frac{13 \nu ^2}{8 \widetilde{r}^9}+\frac{9 \nu }{16 \widetilde{r}^9}\right)\nonumber\\
&~~~~~~~~~~+\widetilde{L}^2 \left(\widetilde{p}_r^2 \left(\frac{69 \nu ^4}{16 \widetilde{r}^7}+\frac{115 \nu ^3}{16 \widetilde{r}^7}-\frac{151 \nu ^2}{16 \widetilde{r}^7}+\frac{3 \nu }{4 \widetilde{r}^7}\right)-\frac{3 \nu ^4}{4 \widetilde{r}^8}-\frac{7 \nu ^3}{2 \widetilde{r}^8}+\frac{107 \nu ^2}{16 \widetilde{r}^8}+\frac{6 \nu }{\widetilde{r}^8}\right)\nonumber\\
&~~~~~~~~~~+\widetilde{p}_r^2 \left(-\frac{2 \nu ^4}{\widetilde{r}^6}-\frac{1925 \nu ^3}{16 \widetilde{r}^6}-\frac{621 \nu ^2}{4 \widetilde{r}^6}+\frac{3 \nu }{\widetilde{r}^6}\right)+\widetilde{p}_r^4 \left(\frac{57 \nu ^4}{8 \widetilde{r}^5}+\frac{341 \nu ^3}{16 \widetilde{r}^5}-\frac{55 \nu ^2}{8 \widetilde{r}^5}+\frac{3 \nu }{16 \widetilde{r}^5}\right)\nonumber\\
&~~~~~~~~~~-\frac{17 \nu ^3}{14 \widetilde{r}^7}+\frac{\left(247616-2547 \pi ^2\right) \nu ^2}{3072 \widetilde{r}^7}+\frac{123 \nu }{4 \widetilde{r}^7}
\Bigg)
\Bigg\}\nonumber\\
&+\left(C^{(0)}_{\rm ES^2}\right)_{(1)}\left((\widetilde{\textbf{S}}_{(1)}\cdot\widetilde{\textbf{r}})(\widetilde{\textbf{S}}_{(1)}\cdot\widetilde{\textbf{L}})\right) \nonumber\\
&~~~~\widetilde{p}_r \Bigg\{
\widetilde{L}^4 \left(\frac{3 \nu ^4}{8 \widetilde{r}^9}-\frac{67 \nu ^3}{16 \widetilde{r}^9}+\frac{39 \nu ^2}{16 \widetilde{r}^9}\right)+\widetilde{L}^2 \left( \left(-\frac{29 \nu ^4}{16 \widetilde{r}^8}+\frac{323 \nu ^3}{16 \widetilde{r}^8}+\frac{16 \nu ^2}{\widetilde{r}^8}\right)+\widetilde{p}_r^2 \left(\frac{9 \nu ^4}{8 \widetilde{r}^7}-\frac{29 \nu ^3}{16 \widetilde{r}^7}-\frac{27 \nu ^2}{8 \widetilde{r}^7}\right)\right)\nonumber\\
&~~~~~~~+\widetilde{p}_r^2 \left(-\frac{93 \nu ^4}{16 \widetilde{r}^6}+\frac{1611 \nu ^3}{16 \widetilde{r}^6}-\frac{72 \nu ^2}{\widetilde{r}^6}\right)+\widetilde{p}_r^4 \left(-\frac{3 \nu ^4}{16 \widetilde{r}^5}+\frac{31 \nu ^3}{2 \widetilde{r}^5}-\frac{93 \nu ^2}{16 \widetilde{r}^5}\right)-\frac{65 \nu ^3}{4 \widetilde{r}^7}-\frac{250693 \nu ^2}{1225 \widetilde{r}^7} \nonumber\\
&~~~~~~~+\frac{1}{q} \bigg(\widetilde{L}^4 \left(-\frac{9 \nu ^4}{16 \widetilde{r}^9}-\frac{37 \nu ^3}{8 \widetilde{r}^9}+\frac{41 \nu ^2}{16 \widetilde{r}^9}-\frac{3 \nu }{16 \widetilde{r}^9}\right)\nonumber\\
&~~~~~~~~~~~~+\widetilde{L}^2 \left( \left(-\frac{29 \nu ^4}{16 \widetilde{r}^8}+\frac{157 \nu ^3}{4 \widetilde{r}^8}+\frac{57 \nu ^2}{16 \widetilde{r}^8}-\frac{3 \nu }{\widetilde{r}^8}\right)+\widetilde{p}_r^2 \left(-\frac{45 \nu ^4}{16 \widetilde{r}^7}-\frac{4 \nu ^3}{\widetilde{r}^7}+\frac{49 \nu ^2}{16 \widetilde{r}^7}-\frac{9 \nu }{8 \widetilde{r}^7}\right)\right)\nonumber\\
&~~~~~~~~~~~~+\widetilde{p}_r^2 \left(-\frac{93 \nu ^4}{16 \widetilde{r}^6}+\frac{2779 \nu ^3}{16 \widetilde{r}^6}+\frac{417 \nu ^2}{16 \widetilde{r}^6}-\frac{9 \nu }{\widetilde{r}^6}\right)+\widetilde{p}_r^4 \left(-\frac{63 \nu ^4}{8 \widetilde{r}^5}+\frac{145 \nu ^3}{16 \widetilde{r}^5}+\frac{19 \nu ^2}{8 \widetilde{r}^5}-\frac{15 \nu }{16 \widetilde{r}^5}\right)\nonumber\\
&~~~~~~~~~~~~-\frac{827 \nu ^3}{28 \widetilde{r}^7}+\frac{\left(2547 \pi ^2-237272\right) \nu ^2}{384 \widetilde{r}^7}-\frac{30 \nu }{\widetilde{r}^7}
\bigg)
\Bigg\}\nonumber\\
&+(1\leftrightarrow 2)
\end{align}
\begin{align}
\widetilde{\mathcal{H}}^{\rm E^2}_{\rm LO}&=
\left(C^{(2)}_{\rm E^2}\right)_{(1)}\left(\widetilde{\textbf{S}}_{(1)}\cdot\widetilde{\textbf{S}}_{(1)}\right)\Bigg\{-\frac{3 \nu ^3}{\widetilde{r}^6}-\frac{1}{q}\frac{3 \nu ^3 }{\widetilde{r}^6}\Bigg\}
+(1\leftrightarrow 2)
\end{align}
\begin{align}
\widetilde{\mathcal{H}}^{\rm E^2S^2}_{\rm LO}&=
\left(C^{(0)}_{\rm E^2S^2}\right)_{(1)} \left(\widetilde{\textbf{S}}_{(1)}\cdot\widetilde{\textbf{S}}_{(1)}\right)\Bigg\{-\frac{\nu ^3}{2 \widetilde{r}^6}-\frac{1}{q}\frac{\nu ^3}{2 \widetilde{r}^6}\Bigg\}\nonumber\\
&+\left(C^{(0)}_{\rm E^2S^2}\right)_{(1)}\left((\widetilde{\textbf{S}}_{(1)}\cdot\widetilde{\textbf{r}})^2\right)\Bigg\{\frac{3 \nu ^3}{2 \widetilde{r}^8}+\frac{1}{q}\frac{3 \nu ^3}{2 \widetilde{r}^8}\Bigg\}
+(1\leftrightarrow 2)
\end{align}

The novel computation of the $\tilde{\mathcal{H}}^{\rm SS}_{\rm N^3LO}$ along with the 
previously known lower order pieces of the Hamiltonian as well as the non-spinning and the spin-orbit Hamiltonian permits us to obtain the total quadratic-in-spin \NNNLO Hamiltonian adopting the EFT formalism.
The known lower order Hamiltonians in the non-spinning and spin-orbit sectors are reported in the appendix C of~\cite{Mandal:2022nty}, 
where as the quadratic-in-spin sector lower order Hamiltonians are reported in the appendix~\ref{app_Ham}.
The total Hamiltonian in the generic frame is also provided in the ancillary file \texttt{Hamiltonian.m} with this article.
Recently, 
the total Hamiltonian at \NNNLO has been also reported in~\cite{Kim:2022bwv}. 
However, as the Hamiltonians are gauge dependent quantities, 
we perform a comparison of the results of gauge-independent observable
in the later part of the article.

\section{Computation of observables with spin}
\label{sec_compputing_observables}
The derived generic Hamiltonian is still gauge dependent, 
because of it's dependence on the radial coordinate. 
So, we can compute observable, which are gauge invariant and comparable with other results in the literature.
In this section, 
we focus on the computation of two gauge invariant observable, namely, the binding energy, and the scattering angle.

For this purpose, we adopt the COM frame, where $\textbf{p}_{(1)}+\textbf{p}_{(2)}=0$, as described in section~\ref{sec_results}. 
We also assume the aligned spin configurations, which implies that the spins are aligned to the direction of the orbital angular momentum of the compact binary.
Such aligned spin configuration is realized by
\begin{align}
\textbf{S}_{(a)}\cdot \textbf{r} = \textbf{S}_{(a)}\cdot \textbf{p} = 0 \implies \textbf{S}_{(a)}\cdot( \textbf{r}\times \textbf{p}) = S_{(a)} L\, ,
\end{align}
with, $L=|\textbf{L}|$ and $S_{(a)}=|\textbf{S}_{(a)}|$.

\subsection{Binding energy for circular orbits with aligned spins}
The gauge invariant relation between the binding energy and the orbital frequency for circular orbits is obtained by eliminating the dependence on the radial coordinate.
For circular orbits we have
\begin{equation}
    \frac{\partial \widetilde{\mathcal{H}}(\widetilde{r},\widetilde{L},\widetilde{S}_{(a)})}{\partial \widetilde{r}}=0\, .
\end{equation}
We invert the above relation to express $\widetilde{r}$ as a function of $\widetilde{L}$. %Using this we can express the Hamiltonian as $\widetilde{H}(\widetilde{L},\widetilde{S}_{(a)})$.
Then we substitute $\widetilde{L}$ as a function of $\widetilde{\omega}$, the orbital frequency defined as
\begin{align}
\widetilde{\omega}=\frac{\partial \widetilde{\mathcal{H}}(\widetilde{L},\widetilde{S}_{(a)})}{\partial \widetilde{L}}\, .
\end{align}
Additionally, we
define a gauge invariant PN parameter $x=\widetilde{\omega}^{2/3}$ .
Following the above procedure with the Hamiltonian given in section \ref{sec_results} we obtain
\begin{align}\label{eq_BE}
E(x,\widetilde{S}_{(a)})= E_{\text{pp}}(x)+E_{\text{SO}}(x,\widetilde{S}_{(a)})+E_{\text{SS}}(x,\widetilde{S}_{(a)})\, ,
\end{align} 
where $E_{\text{pp}}$ and $E_{\text{SO}}$ are reported in \cite{Mandal:2022nty}, and
\begin{align}\label{eq_BE}
E_{\text{SS}}(x,\widetilde{S}_{(a)}) &= E_{\rm S1S2}(x,\widetilde{S}_{(a)}) + E_{\rm S^2}(x,\widetilde{S}_{(a)}) \nonumber\\
&+ E_{\rm ES^2}(x,\widetilde{S}_{(a)}) + E_{\rm E^2S^2}(x,\widetilde{S}_{(a)}) + E_{\rm E^2}(x,\widetilde{S}_{(a)})\, .
\end{align}
The individual terms in the above equation are given as, 
\begin{align}
\label{eq:BE_S1S2_N3LO}
E_{\rm S1S2}(x,\widetilde{S})&= \widetilde{S}_{(1)}\widetilde{S}_{(2)} \Bigg\{
x^3\bigg\{\nu\bigg\}+x^4\bigg\{\frac{5}{6}\nu+\frac{5}{18}\nu^2\bigg\}+x^5\bigg\{\frac{35}{8}\nu-\frac{1001}{72}\nu^2-\frac{371}{216}\nu^3\bigg\}\nonumber\\
&~~~~~~~~~~~~~~~~~~~~+x^6\bigg\{\frac{243}{16}\nu-\left(\frac{2107}{16}-\frac{123}{32}\pi^2\right)\nu^2+\frac{147}{8}\nu^3+\frac{13}{16}\nu^4\bigg\}
\Bigg\} \, ,
\end{align}
\begin{align}
\label{eq:BE_S2_N3LO}
E_{\rm S^2}(x,\widetilde{S})&= \widetilde{S}_{(1)}^2 \Bigg\{
x^4\bigg\{\frac{25}{18}\nu^2+\frac{1}{q}\left(-\frac{5}{2}\nu+\frac{5}{6}\nu^2\right)\bigg\}\nonumber\\
&~~~~~~~~~+x^5\bigg\{\frac{10}{3}\nu^2-\frac{749}{108}\nu^3+\frac{1}{q}\left(-\frac{21}{4}\nu-\frac{7}{6}\nu^2-\frac{217}{36}\nu^3\right)\bigg\}\nonumber\\
&~~~~~~~~~+x^6\bigg\{\frac{1947}{112}\nu^2-\frac{48357 }{560}\nu ^3+\frac{159 }{16}\nu ^4\nonumber\\
&~~~~~~~~~~~~~~~~~+\frac{1}{q}\left(-\frac{243  }{16}\nu+\left(\frac{747}{16}-\frac{189 \pi ^2}{2048}\right) \nu^2 -\frac{13731 }{280}\nu ^3+\frac{153 }{16}\nu ^4\right)\bigg\}
\Bigg\} \, \nonumber\\
& + (1\leftrightarrow 2) \, ,
\end{align}
\begin{align}
\label{eq:BE_ES2_N3LO}
E_{\rm ES^2}(x,\widetilde{S})&=\left(C^{(0)}_{\rm ES^2}\right)_{(1)} \widetilde{S}_{(1)}^2 \Bigg\{x^3 \bigg\{
\frac{1}{q}\frac{\nu}{2}\bigg\}+x^4\bigg\{\frac{5 }{3}\nu^2+\frac{1}{q}\left(\frac{5 }{4}\nu+\frac{5}{4} \nu^2\right) \bigg\}\nonumber\\
&~~~~~~~~~~~~~~~~~+x^5\bigg\{\frac{31 }{4}\nu^2 -\frac{35 }{18}\nu ^3+\frac{1}{q}\left(\frac{63 }{16}\nu+\frac{77 }{48}\nu^2 -\frac{91 }{48}\nu ^3\right)\bigg\}\nonumber\\
&~~~~~~~~~~~~~~~~~+x^6\bigg\{\frac{789 }{28}\nu^2 -\frac{156 }{7}\nu ^3+\frac{5 }{8}\nu ^4\nonumber\\
&~~~~~~~~~~~~~~~~~~~~~~~~~+\frac{1}{q}\left(\frac{405}{32}\nu+\left(\frac{3747 \pi ^2}{2048}-\frac{2389}{32}\right) \nu^2 -\frac{555 }{56}\nu ^3+\frac{21 }{32}\nu ^4\right)\bigg\}
\Bigg\} \, \nonumber\\
& + (1\leftrightarrow 2) \, ,
\end{align}
\begin{align}
\label{eq:BE_E2_N3LO}
E_{\rm E^2}(x,\widetilde{S})&=\left(C^{(2)}_{\rm E^2}\right)_{(1)} \widetilde{S}_{(1)}^2 x^6 \bigg\{ 9\nu^3\left(
1+\frac{1}{q}\right)\bigg\} + (1\leftrightarrow 2) \, ,
\end{align}
\begin{align}
\label{eq:BE_E2S2_N3LO}
E_{\rm E^2S^2}(x,\widetilde{S})&=\left(C^{(0)}_{\rm E^2S^2}\right)_{(1)} \widetilde{S}_{(1)}^2 x^6 \bigg\{ \frac{3\nu^3}{2}\left(
1+\frac{1}{q}\right)\bigg\} + (1\leftrightarrow 2) \, .
\end{align}
The equation \eqref{eq:BE_S1S2_N3LO} agrees with previously known classical results of \cite{Antonelli:2020ybz}, and the equations \eqref{eq:BE_S1S2_N3LO}, \eqref{eq:BE_S2_N3LO}, \eqref{eq:BE_ES2_N3LO} and \eqref{eq:BE_E2S2_N3LO} agree with recent results given in \cite{Kim:2021rfj} derived using similar EFT techniques as presented in this paper.

\subsection{Scattering angle with aligned spins}
In this section, we compute the scattering angle considering an aligned spin binary system following \cite{Vines:2018gqi}. 
First, we re-scale the spin variables as $a_{(a)}=S_{(a)}/(m_{(a)}c)$.
In the COM, the Hamiltonian $\mathcal{H}$ is expressed as a function of $\widetilde{p}_r$, $L$, $r$, and $S{(a)}$ and inverting that we obtain $\widetilde{p}_r(\mathcal{H},L,r,S_{(a)})$.
Then the scattering angle $\chi$ is given by 
\begin{align}\label{eq_scattering_angle_1}
\chi(\mathcal{H},L,S_{(a)})=-\int dr \frac{\partial \widetilde{p}_r(\mathcal{H},L,r,S_{(a)})}{\partial L}  - \pi\, .
\end{align}
We invert the relation between the Lorentz factor $\gamma$ and the the total energy per 
total rest mass $\Gamma = \mathcal{H}/(Mc^2)$ given by 
\begin{align}\label{eq_gamma_for_v}
\gamma=\frac{1}{\sqrt{1-v^2/c^2}}=1+\frac{\Gamma^2-1}{2\nu}\, ,
\end{align}
where, $v\equiv|\dot{\textbf{r}}|$ is the relative velocity of the compact objects, and the relation between the total angular momentum $L$ and the impact parameter $b$ given by 
\begin{align}\label{eq_L_for_b}
L=\frac{\mu\gamma v b}{\Gamma} + Mc \left(\frac{\Gamma-1}{2}\right) \left(a_+ - \frac{\delta}{\Gamma} a_-\right)\, ,
\end{align}
where, $a_{(+)}=a_{(1)}+a_{(2)}$ and $a_{(-)}=a_{(1)}-a_{(2)}$. 
With the above inversions, we trade $H$ for $v$
and $L$ for $b$.
This allows us to express the scattering angle as
\begin{align}
\chi(v,b,S_{(a)})=-\frac{\gamma}{\mu\gamma v}\int dr \frac{\partial \widetilde{p}_r(v,b,r,S_{(a)})}{\partial b}  - \pi\, .
\end{align}
Now, applying the above procedure with the Hamiltonian given in section \ref{sec_results}, we obtain the scattering angle computed in the COM for aligned spins, which can be expressed as
\begin{align}\label{eq_SA}
\chi(v,b,S_{(a)})= \chi_{\text{pp}}(v,b)+\chi_{\text{SO}}(v,b,S_{(a)}) + \chi_{\rm SS}(v,b,S_{(a)}) 
\end{align} 
where $\chi_{\text{pp}}$ and $\chi_{\text{SO}}$ are reported in \cite{Mandal:2022nty}, and
\begin{align}
    \chi_{\rm SS}(v,b,S_{(a)})&= \chi_{\rm S1S2}(v,b,S_{(a)}) + \chi_{\rm S^2}(v,b,S_{(a)}) \nonumber\\
    &+ \chi_{\rm ES^2}(v,b,S_{(a)}) + \chi_{\rm E^2S^2}(v,b,S_{(a)}) + \chi_{\rm E^2}(v,b,S_{(a)}) ~.
\end{align}
The individual terms in the above equation are given as, 
\begin{align}
\label{eq:SA_S1S2_N3LO}
\frac{\chi_{\rm S1S2}+\chi_{\rm S^2}}{\Gamma}=  \frac{1}{b^2c^2} 
&\begin{bmatrix}%
a_{(+)}^2  & ~ \delta a_{(+)} a_{(-)} & ~ a_{(-)}^2
\end{bmatrix}
\cdot\Bigg(
\left(\frac{G_NM}{v^2b}\right)\Bigg\{  \begin{bmatrix}1 \\ 0 \\-1\end{bmatrix} + \begin{bmatrix}1 \\ 0 \\-1\end{bmatrix} \left(\frac{v^2}{c^2}\right) \Bigg\}\nonumber\\
&+\pi \left(\frac{G_NM}{v^2b}\right)^2 \Bigg\{ \frac{3}{4}  \begin{bmatrix}1 \\ 0\\ -1\end{bmatrix} + \frac{3}{16}  \begin{bmatrix}41 \\ 10\\ -19\end{bmatrix} \left(\frac{v^2}{c^2}\right) + \frac{3}{128}  \begin{bmatrix} 55\\ -10\\ -41\end{bmatrix} \left(\frac{v^4}{c^4}\right) \Bigg\}\nonumber\\
&+ \left(\frac{G_NM}{v^2b}\right)^3 \Bigg\{ 2  \begin{bmatrix}1 \\ 0\\-1\end{bmatrix}+  \begin{bmatrix}102-2\nu \\ 32\\ -38-6\nu\end{bmatrix} \left(\frac{v^2}{c^2}\right) + \frac{1}{14} \begin{bmatrix}2332-499\nu \\ 688 \\ -748-37\nu\end{bmatrix} \left(\frac{v^4}{c^4}\right)\nonumber\\
&\quad\quad\quad\quad\quad\quad\quad+ \frac{1}{140} \begin{bmatrix}1704-9925\nu \\ -288\\-1096+5377\nu\end{bmatrix} \left(\frac{v^6}{c^6}\right)\Bigg\}\nonumber\\
&+ \pi\left(\frac{G_NM}{v^2b}\right)^4 \Bigg\{ \frac{15}{16}  \begin{bmatrix}63-2\nu \\ 22 \\ -21-6\nu\end{bmatrix}\left(\frac{v^2}{c^2}\right) -\frac{15}{448}  \begin{bmatrix}-11063+2638\nu \\ -4206+308\nu \\ 2657+790\nu\end{bmatrix} \left(\frac{v^4}{c^4}\right) \nonumber\\
&\quad\quad\quad\quad\quad\quad\quad-\frac{3}{229376}  \begin{bmatrix} 19102720+(-29293696+135555\pi^2) \\ -256(-24640+11507\nu) \\ -4730880+(8911744-139965\pi^2)\nu  \end{bmatrix} \left(\frac{v^6}{c^6}\right)\Bigg\}\Bigg)\nonumber\\
&+\mathcal{O}\left(G_N^5,\frac{v^8}{c^8}\right) \, ,
\end{align}
\begin{align}
\label{eq:SA_ES2_N3LO}
\frac{\chi_{\rm ES^2}}{\Gamma}=  \frac{1}{b^2c^2} 
&\begin{bmatrix}%
 a_{\rm ES^2 (+)}^2 & ~\delta a_{\rm ES^2(-)}^2
\end{bmatrix}
\cdot \Bigg\{  
\left(\frac{G_NM}{v^2b}\right)\bigg\{   \begin{bmatrix} 2\\ 0\end{bmatrix} + \begin{bmatrix} 2\\ 0\end{bmatrix} \left(\frac{v^2}{c^2}\right) \bigg\}\nonumber\\
&+\pi \left(\frac{G_NM}{v^2b}\right)^2 \bigg\{  \frac{1}{2}\begin{bmatrix} 3\\ 0\end{bmatrix} + \frac{1}{4}\begin{bmatrix} 27\\ 3\end{bmatrix}  \left(\frac{v^2}{c^2}\right) + \frac{1}{64}\begin{bmatrix} 117\\ 57\end{bmatrix}  \left(\frac{v^4}{c^4}\right) \bigg\}\nonumber\\
&+ \left(\frac{G_NM}{v^2b}\right)^3 \bigg\{ \begin{bmatrix} 4\\ 0\end{bmatrix} +   \begin{bmatrix} 76-4\nu\\ 8\end{bmatrix}  \left(\frac{v^2}{c^2}\right) + \frac{1}{7}\begin{bmatrix} 748-131\nu\\ 216\end{bmatrix} \left(\frac{v^4}{c^4}\right) + \frac{1}{70}\begin{bmatrix} 1096-5013\nu\\ 704\end{bmatrix}  \left(\frac{v^6}{c^6}\right)\bigg\}\nonumber\\
&+ \pi\left(\frac{G_NM}{v^2b}\right)^4 \bigg\{ \frac{15}{4}\begin{bmatrix} 11-\nu\\ 1\end{bmatrix}  \left(\frac{v^2}{c^2}\right) + \frac{15}{224}\begin{bmatrix} 2867-708\nu\\ 627-28\nu\end{bmatrix}  \left(\frac{v^4}{c^4}\right) \nonumber\\
&\quad\quad\quad+ \frac{15}{114688}\begin{bmatrix}1026816-(2184832+27111\pi^2)\nu\\ 128(3262-395\nu)\end{bmatrix}  \left(\frac{v^6}{c^6}\right)\bigg\} \Bigg\}
+\mathcal{O}\left(G_N^5,\frac{v^8}{c^8}\right)\, ,
\end{align}
\begin{align}
\label{eq:SA_E2_N3LO}
\frac{\chi_{\rm E^2}}{\Gamma}=  \frac{1}{b^2c^2} 
&\begin{bmatrix}%
 a_{\rm E^2(+)}^2 & ~\delta a_{\rm E^2(-)}^2
\end{bmatrix}
\cdot \Bigg\{ 
 \pi\left(\frac{G_NM}{v^2b}\right)^4  \frac{45}{16}\nu \begin{bmatrix}1\\ 1\end{bmatrix}  \left(\frac{v^6}{c^6}\right)  \Bigg\}
+\mathcal{O}\left(G_N^5,\frac{v^8}{c^8}\right)\, ,
\end{align}
\begin{align}
\label{eq:SA_E2S2_N3LO}
\frac{\chi_{\rm E^2S^2}}{\Gamma}=  \frac{1}{b^2c^2} 
&\begin{bmatrix}%
 a_{\rm E^2S^2 (+)}^2 & ~\delta a_{\rm E^2S^2(-)}^2
\end{bmatrix}
\cdot \Bigg\{ 
 \pi\left(\frac{G_NM}{v^2b}\right)^4  \frac{15}{32}\nu \begin{bmatrix}1\\ 1\end{bmatrix}  \left(\frac{v^6}{c^6}\right)  \Bigg\}
+\mathcal{O}\left(G_N^5,\frac{v^8}{c^8}\right)\, .
\end{align}
where, $a_{\rm ES^2(\pm)}^2 = \left(C^{(0)}_{\rm ES^2}\right)_{(1)} a_{(1)}^2~ \pm \left(C^{(0)}_{\rm ES^2}\right)_{(2)} a_{(2)}^2$
%   a_{\rm E^2S^2(\pm)}^2 &= \left(C^{(0)}_{\rm E^2S^2}\right)_{(1)} a_{(1)}^2 \pm \left(C^{(0)}_{\rm ES^2}\right)_{(2)} a_{(2)}^2\\
%   a_{\rm E^2(\pm)}^2 &= ~ \left(C^{(2)}_{\rm E^2}\right)_{(1)} a_{(1)}^2~ \pm ~\left(C^{(2)}_{\rm E^2}\right)_{(2)} a_{(2)}^2
and simillarly for $a_{\rm E^2S^2(\pm)}^2$ and $a_{\rm E^2(\pm)}^2$.
The equation \eqref{eq:SA_S1S2_N3LO} agrees with previously known classical results of \cite{Antonelli:2020ybz}, and the equations \eqref{eq:SA_S1S2_N3LO}, \eqref{eq:SA_ES2_N3LO} and \eqref{eq:SA_E2S2_N3LO} agree with recent results given in \cite{Kim:2022bwv} derived using similar EFT techniques as presented in this paper.
Note that the scattering angle depends on the same combination of $C^{(2)}_{\rm E^2}$ and $C^{(0)}_{\rm E^2S^2}$ as the binding energy.
Matching both constants independently hence requires another observable in a more asymmetric (e.g. precessing) kinematic regime.

\section{Conclusion}\label{sec_Conclusion}

In this work, we presented the complete evaluation of \NNNLO Post-Newtonian correction to 
the quadratic-in-spin Hamiltonian for the spinning compact binaries, 
within the EFT Feynman diagrammatic approach of GR.
Together with our earlier study on the spin-orbit Hamiltonian \cite{Mandal:2022nty},
this is at the current state-of-the-art result for the conservative part of the spinning sectors up to quadratic-in-spin.

The necessary Feynman diagrams in momentum space were generated using Feynman rules derived from the EFT Lagrangian describing the spinning compact objects.
The corresponding Feynman amplitudes were written as a linear combination of master integrals, employing the integration-by-parts identities for dimensionally regularized integrals. 
The contribution of each diagram to the effective potential was obtained upon Fourier transform, and Laurent series expanded around $d=3+\epsilon$ space dimensions.
Finally, 
the Hamiltonian was derived by applying Legendre transform, and canonical transformations were employed to remove the non-physical divergences and spurious logarithmic terms.
Furthermore, we computed two specific gauge invariant observable, namely, the binding energy with aligned spins in the case of circular orbit, and the scattering angle for the case of  aligned spins. 
These results were found in agreement with the corresponding expressions available in literature, previously obtained using the self-force formalism as well as the EFT framework.

The obtained results depend on three undetermined Wilson coefficients, 
namely,
$C^{(2)}_{\rm E^2}$, 
$C^{(0)}_{\rm ES^2}$, 
and
$C^{(0)}_{\rm E^2S^2}$. 
They arise from the non-minimal couplings at linear and quadratic order in curvature in the EFT Lagrangian of the point particles. The coefficient $C^{(0)}_{\rm ES^2}$ starts contributing from 2PN (LO $\rm S^2$ sector), whereas the other two coefficients $C^{(2)}_{\rm E^2}$ and $C^{(0)}_{\rm E^2S^2}$ contribute for the first time at 5PN.
Out of them, $C^{(0)}_{\rm ES^2}$ is related to the spin-induced quadruple moment of the compact object:
for Kerr black holes, it is known to be 1 \cite{DEath:1975wqz, RevModPhys.52.299}; 
and for neutron stars its value ranges within 2-8 \cite{Laarakkers:1997hb}.
The other two coefficients, $C^{(2)}_{\rm E^2}$ and $C^{(0)}_{\rm E^2S^2}$, encode quadrupolar deformations, due to an external field and spin-square effects.
Their determination requires the matching against stationary linear perturbations of spinning compact objects -- see~\cite{Poisson:2014gka, Pani:2015hfa, LeTiec:2020spy, Ivanov:2022qqt} for crucial work in that direction. 
The observable presented in this work could also be used to match a combination of $C^{(2)}_{\rm E^2}$ and $C^{(0)}_{\rm E^2S^2}$ to the motion of a small body within the self-force formalism.
The dependence of gravitational waves from compact binaries on $C^{(2)}_{\rm E^2}$ and $C^{(0)}_{\rm E^2S^2}$ provides a probe of the nature of black holes and encodes information about the equation of state of neutron stars.

\subsection*{Acknowledgements}
The work of M.K.M is supported by Fellini - Fellowship for Innovation at INFN funded by the European Union's Horizon 2020 research and innovation programme under the Marie Sk{\l}odowska-Curie grant agreement No 754496. 
RP is grateful to IISER Bhopal for the fellowship. RP's research is funded by the Deutsche Forschungsgemeinschaft (DFG, German Research Foundation), Projektnummer 417533893/GRK2575 “Rethinking Quantum Field Theory”.  
\appendix
\section{Notation and convention} \label{app_notation_and_convention}

\begin{subequations}   
	\begin{eqnarray} 
	\textrm{Spacetime metric}\quad \quad && \eta_{\mu \nu }=(1,-1,-1,-1)\\
	\textrm{4 dimensional indices}\quad \quad && \mu,\nu \\
	\textrm{3 dimensional indices}\quad \quad && i,j 
	\\
	\textrm{Compact object label}\quad \quad&& _{(a)} \quad \textrm{where } a=\{1,2\}\\
	\textrm{Time derivative}\quad \quad&& \dot{}\\
	\textrm{Position of $a^{\text{th}}$ object}\quad \quad&&\textbf{x}_{(a)}\\
	\textrm{Velocity of $a^{\text{th}}$ object}\quad \quad&&\textbf{v}_{(a)}\equiv\dot{\textbf{x}}_{(a)}\\
	\textrm{Acceleration of $a^{\text{th}}$ object}\quad \quad&&\textbf{a}_{(a)}\equiv\ddot{\textbf{x}}_{(a)}\\
	\textrm{Separation vector for binary}\quad \quad&&\textbf{r}\equiv\textbf{x}_{(a)}-\textbf{x}_{(a)}\\
	\textrm{Separation distance for binary}\quad \quad&&r\equiv|\textbf{r}|\\
	\textrm{Separation unit vector for binary}\quad \quad&&\textbf{n}\equiv\frac{\textbf{r}}{r}\\
	\textrm{Angular momentum of the binary}\quad \quad&& \textbf{L}\equiv(\textbf{r}\times \textbf{p})\\
    \textrm{Spin vector of $a^{\text{th}}$ object}\quad \quad&& \textbf{S}_{(a)}^i\equiv\bm{\epsilon}^{ijk} \textbf{S}_{(a)}^{jk}\\
	\int_p\quad \quad&& \int \frac{d^dp}{(2\pi)^d}\\
	\int_\textbf{p}\quad \quad&& \int \frac{d^3p}{(2\pi)^3}\\
	\textrm{Center of mass coordinates}\quad \quad&&\textbf{p}_{(1)}+\textbf{p}_{(2)}=0\\
	\textrm{Circular orbits}\quad \quad&&\widetilde{p}_r\equiv\textbf{p}\cdot \textbf{n}=0 \quad\quad\textrm{and}\quad\quad\dot{p}_r=0\\
	\textrm{Aligned spins}\quad \quad&&\textbf{S}_{(a)}\cdot \textbf{r} =\textbf{S}_{(a)}\cdot \textbf{p} = 0\\
	\textrm{Exchange of particle label }(1\leftrightarrow 2) \quad \quad&& \left(_{(1)}\leftrightarrow _{(2)}\right), \left(q\leftrightarrow\frac{1}{q}\right), \left(\delta \leftrightarrow -\delta\right)
	\end{eqnarray} 
\end{subequations}

\section{Lower-order Hamiltonians}\label{app_Ham}
In this appendix we give the results for all the lower order Hamiltonians given in eq.(\ref{eq_Ham_S1S2}).

\subsection{Spin1-Spin2 sector up to \NNLO}

\begin{align}
\widetilde{\mathcal{H}}^{\rm S1S2}_{\text{LO}}= 
\left(\widetilde{\textbf{S}}_{(1)}\cdot\widetilde{\textbf{S}}_{(2)}\right)\left\{
-\frac{\nu}{\widetilde{r}^3}
\right\}
+\left((\widetilde{\textbf{S}}_{(1)}\cdot\widetilde{r})(\widetilde{\textbf{S}}_{(2)}\cdot\widetilde{r})\right)\left\{
\frac{3 \nu}{\widetilde{r}^5}
\right\}
\end{align}

\begin{align}
\widetilde{\mathcal{H}}^{\rm S1S2}_{\text{NLO}}&= 
\left(\widetilde{\textbf{S}}_{(1)}\cdot\widetilde{\textbf{S}}_{(2)}\right)\left\{
\widetilde{L}^2 \left(\frac{\nu }{\widetilde{r}^5}-\frac{\nu ^2}{4 \widetilde{r}^5}\right)+\widetilde{p}_r^2 \left(-\frac{7 \nu ^2}{4 \widetilde{r}^3}-\frac{\nu }{2 \widetilde{r}^3}\right)+\frac{5 \nu }{\widetilde{r}^4}
\right\}\nonumber\\
&+\left((\widetilde{\textbf{S}}_{(1)}\cdot\widetilde{\textbf{r}})(\widetilde{\textbf{S}}_{(2)}\cdot\widetilde{\textbf{r}})\right)\left\{
\frac{3 \widetilde{L}^2 \nu ^2}{2 \widetilde{r}^7}+\widetilde{p}_r^2 \left(\frac{23 \nu ^2}{4 \widetilde{r}^5}+\frac{\nu }{2 \widetilde{r}^5}\right)-\frac{11 \nu }{\widetilde{r}^6}
\right\}\nonumber\\
&+\left((\widetilde{\textbf{S}}_{(1)}\cdot\widetilde{\textbf{L}})(\widetilde{\textbf{S}}_{(2)}\cdot\widetilde{\textbf{L}})\right)\left\{
-\frac{\nu ^2}{\widetilde{r}^5}-\frac{5 \nu }{2 \widetilde{r}^5}
\right\}\nonumber\\
&+\left((\widetilde{\textbf{S}}_{(1)}\cdot\widetilde{\textbf{r}})(\widetilde{\textbf{S}}_{(2)}\cdot\widetilde{\textbf{L}})\right) \widetilde{p}_r \left\{
\frac{1}{q}\frac{3 \nu ^2}{2 \widetilde{r}^5}+ \left(-\frac{\nu ^2}{4 \widetilde{r}^5}-\frac{\nu }{\widetilde{r}^5}\right)
\right\}\nonumber\\
&+\left((\widetilde{\textbf{S}}_{(1)}\cdot\widetilde{\textbf{L}})(\widetilde{\textbf{S}}_{(2)}\cdot\widetilde{\textbf{r}})\right) \widetilde{p}_r \left\{
\left(\frac{\nu }{2 \widetilde{r}^5}-\frac{13 \nu ^2}{4 \widetilde{r}^5}\right)-\frac{1}{q} \frac{3 \nu ^2 }{2 \widetilde{r}^5}
\right\}
\end{align}

\begin{align}
\widetilde{\mathcal{H}}^{\rm S1S2}_{\rm N^2LO}&= 
\left(\widetilde{\textbf{S}}_{(1)}\cdot\widetilde{\textbf{S}}_{(2)}\right)\Bigg\{
\widetilde{L}^4 \left(-\frac{\nu ^3}{8 \widetilde{r}^7}+\frac{9 \nu ^2}{8 \widetilde{r}^7}-\frac{3 \nu }{4 \widetilde{r}^7}\right)+\widetilde{L}^2 \left(\widetilde{p}_r^2 \left(-\frac{5 \nu ^3}{8 \widetilde{r}^5}+\frac{237 \nu ^2}{16 \widetilde{r}^5}-\frac{3 \nu }{8 \widetilde{r}^5}\right)+\frac{61 \nu ^2}{16 \widetilde{r}^6}-\frac{19 \nu }{4 \widetilde{r}^6}\right) \nonumber\\
&~~~~~~~~~~~~~~~~~~~~ +\widetilde{p}_r^2 \left(\frac{75 \nu ^2}{16 \widetilde{r}^4}+\frac{9 \nu }{4 \widetilde{r}^4}\right)+\widetilde{p}_r^4 \left(-\frac{19 \nu ^3}{8 \widetilde{r}^3}-\frac{81 \nu ^2}{16 \widetilde{r}^3}+\frac{3 \nu }{8 \widetilde{r}^3}\right)-\frac{31 \nu ^2}{8 \widetilde{r}^5}-\frac{27 \nu }{2 \widetilde{r}^5} 
\Bigg\}\nonumber\\
&+\left((\widetilde{\textbf{S}}_{(1)}\cdot\widetilde{\textbf{r}})(\widetilde{\textbf{S}}_{(2)}\cdot\widetilde{\textbf{r}})\right)\Bigg\{
\widetilde{L}^4 \left(\frac{9 \nu ^3}{8 \widetilde{r}^9}-\frac{3 \nu ^2}{4 \widetilde{r}^9}\right)+\widetilde{L}^2 \left(\widetilde{p}_r^2 \left(\frac{7 \nu ^3}{2 \widetilde{r}^7}-\frac{97 \nu ^2}{8 \widetilde{r}^7}-\frac{3 \nu }{8 \widetilde{r}^7}\right)-\frac{17 \nu ^2}{4 \widetilde{r}^8}-\frac{9 \nu }{4 \widetilde{r}^8}\right) \nonumber\\
&~~~~~~~~~~~~~~~~~~~~~~~~~~~~~~ +\widetilde{p}_r^2 \left(\frac{3 \nu }{4 \widetilde{r}^6}-\frac{723 \nu ^2}{16 \widetilde{r}^6}\right)+\widetilde{p}_r^4 \left(\frac{67 \nu ^3}{8 \widetilde{r}^5}+\frac{49 \nu ^2}{16 \widetilde{r}^5}-\frac{3 \nu }{8 \widetilde{r}^5}\right)+\frac{95 \nu ^2}{8 \widetilde{r}^7}+\frac{53 \nu }{2 \widetilde{r}^7} 
\Bigg\}\nonumber\\
&+\left((\widetilde{\textbf{S}}_{(1)}\cdot\widetilde{\textbf{L}})(\widetilde{\textbf{S}}_{(2)}\cdot\widetilde{\textbf{L}})\right)\Bigg\{
\widetilde{L}^2 \left(-\frac{\nu ^3}{\widetilde{r}^7}-\frac{\nu ^2}{2 \widetilde{r}^7}+\frac{15 \nu }{8 \widetilde{r}^7}\right)+\widetilde{p}_r^2 \left(-\frac{5 \nu ^3}{2 \widetilde{r}^5}-\frac{181 \nu ^2}{8 \widetilde{r}^5}+\frac{15 \nu }{8 \widetilde{r}^5}\right)+\frac{5 \nu ^2}{2 \widetilde{r}^6}+\frac{27 \nu }{2 \widetilde{r}^6} 
\Bigg\}\nonumber\\
&+\left((\widetilde{\textbf{S}}_{(1)}\cdot\widetilde{\textbf{r}})(\widetilde{\textbf{S}}_{(2)}\cdot\widetilde{\textbf{L}})\right) \widetilde{p}_r\Bigg\{
\widetilde{L}^2  \left(-\frac{5 \nu ^3}{8 \widetilde{r}^7}-\frac{43 \nu ^2}{8 \widetilde{r}^7}+\frac{3 \nu }{4 \widetilde{r}^7}\right)+ \left(-\frac{3 \nu ^3}{2 \widetilde{r}^6}+\frac{175 \nu ^2}{8 \widetilde{r}^6}+\frac{2 \nu }{\widetilde{r}^6}\right)\nonumber\\
&~~~~~~~~~~~~~~~~~~~~~~~~~~~~~~~~~+\widetilde{p}_r^2 \left(-\frac{\nu ^3}{4 \widetilde{r}^5}-\frac{35 \nu ^2}{16 \widetilde{r}^5}+\frac{3 \nu }{4 \widetilde{r}^5}\right) \nonumber\\
&~~~~~~~~~~~~~~~~~~~~~~~~~~~~~~~~~ +\frac{1}{q} \left(\widetilde{L}^2  \left(\frac{3 \nu ^3}{4 \widetilde{r}^7}-\frac{9 \nu ^2}{8 \widetilde{r}^7}\right)+\left(-\frac{3 \nu ^3}{2 \widetilde{r}^6}-\frac{7 \nu ^2}{\widetilde{r}^6}\right)+\widetilde{p}_r^2 \left(\frac{9 \nu ^3}{2 \widetilde{r}^5}-\frac{9 \nu ^2}{8 \widetilde{r}^5}\right)\right)
\Bigg\}\nonumber\\
&+\left((\widetilde{\textbf{S}}_{(1)}\cdot\widetilde{L})(\widetilde{\textbf{S}}_{(2)}\cdot\widetilde{\textbf{r}})\right) \widetilde{p}_r \Bigg\{
\widetilde{L}^2  \left(-\frac{17 \nu ^3}{8 \widetilde{r}^7}-\frac{19 \nu ^2}{8 \widetilde{r}^7}-\frac{3 \nu }{8 \widetilde{r}^7}\right)+ \left(\frac{3 \nu ^3}{2 \widetilde{r}^6}+\frac{275 \nu ^2}{8 \widetilde{r}^6}-\frac{5 \nu }{\widetilde{r}^6}\right)\nonumber\\
&~~~~~~~~~~~~~~~~~~~~~~~~~~~~~~~~~+\widetilde{p}_r^2 \left(-\frac{37 \nu ^3}{4 \widetilde{r}^5}+\frac{73 \nu ^2}{16 \widetilde{r}^5}-\frac{3 \nu }{8 \widetilde{r}^5}\right) \nonumber\\
&~~~~~~~~~~~~~~~~~~~~~~~~~~~~~~~~~ +\frac{1}{q} \left(\widetilde{L}^2 \left(\frac{9 \nu ^2}{8 \widetilde{r}^7}-\frac{3 \nu ^3}{4 \widetilde{r}^7}\right)+ \left(\frac{3 \nu ^3}{2 \widetilde{r}^6}+\frac{7 \nu ^2}{\widetilde{r}^6}\right)+\widetilde{p}_r^2 \left(\frac{9 \nu ^2}{8 \widetilde{r}^5}-\frac{9 \nu ^3}{2 \widetilde{r}^5}\right)\right)
\Bigg\}
\end{align}

\subsection{Spin1-Spin1 and Spin2-Spin2 sector up to \NNLO}

\begin{align}
\widetilde{\mathcal{H}}^{\rm S^2}_{\text{LO}}&= 0
\end{align}

\begin{align}
\widetilde{\mathcal{H}}^{\rm S^2}_{\text{NLO}}&= 
\left(\widetilde{\textbf{S}}_{(1)}\cdot\widetilde{\textbf{S}}_{(1)}\right)\left\{
\frac{3 \widetilde{L}^2 \nu ^2}{2 \widetilde{r}^5}-\frac{3 \nu ^2 \widetilde{p}_r^2}{8 \widetilde{r}^3}+\frac{7 \nu ^2}{8 \widetilde{r}^4}+ \frac{1}{q} \left(\widetilde{L}^2 \left(\frac{11 \nu ^2}{8 \widetilde{r}^5}-\frac{3 \nu }{2 \widetilde{r}^5}\right)+\widetilde{p}_r^2 \left(\frac{3 \nu }{8 \widetilde{r}^3}-\frac{\nu ^2}{2 \widetilde{r}^3}\right)+\frac{7 \nu ^2}{8 \widetilde{r}^4}+\frac{9 \nu }{8 \widetilde{r}^4}\right)
\right\}\nonumber\\
&+\left((\widetilde{\textbf{S}}_{(1)}\cdot\widetilde{\textbf{r}})^2\right)\left\{
+\frac{3 \nu ^2 \widetilde{p}_r^2}{8 \widetilde{r}^5}-\frac{15 \nu ^2}{8 \widetilde{r}^6}+\frac{1}{q} \left(\widetilde{p}_r^2 \left(\frac{\nu ^2}{2 \widetilde{r}^5}-\frac{3 \nu }{8 \widetilde{r}^5}\right)-\frac{15 \nu ^2}{8 \widetilde{r}^6}-\frac{9 \nu }{8 \widetilde{r}^6}\right)
\right\}\nonumber\\
&+\left((\widetilde{\textbf{S}}_{(1)}\cdot\widetilde{\textbf{L}})^2\right)\left\{
-\frac{33 \nu ^2}{8 \widetilde{r}^5}+\frac{1}{q} \left(\frac{33 \nu }{8 \widetilde{r}^5}-\frac{29 \nu ^2}{8 \widetilde{r}^5}\right)
\right\}\nonumber\\
&+\left((\widetilde{\textbf{S}}_{(1)}\cdot\widetilde{\textbf{r}})(\widetilde{\textbf{S}}_{(1)}\cdot\widetilde{\textbf{L}})\right) \widetilde{p}_r \left\{
-\frac{9 \nu ^2}{8 \widetilde{r}^5}+\frac{1}{q} \left(\frac{9 \nu }{8 \widetilde{r}^5}-\frac{7 \nu ^2}{8 \widetilde{r}^5}\right)
\right\}\nonumber\\
&+(1\leftrightarrow 2)
\end{align}

\begin{align}
\widetilde{\mathcal{H}}^{\rm S^2}_{\rm N^2LO}&= 
\left(\widetilde{\textbf{S}}_{(1)}\cdot\widetilde{\textbf{S}}_{(1)}\right)\Bigg\{
\widetilde{L}^4 \left(\frac{15 \nu ^3}{8 \widetilde{r}^7}-\frac{7 \nu ^2}{8 \widetilde{r}^7}\right)+\widetilde{L}^2 \left(\widetilde{p}_r^2 \left(\frac{87 \nu ^3}{16 \widetilde{r}^5}-\frac{\nu ^2}{4 \widetilde{r}^5}\right)-\frac{13 \nu ^3}{16 \widetilde{r}^6}-\frac{35 \nu ^2}{4 \widetilde{r}^6}\right) \nonumber\\
&~~~~~~~~~~~~~~~~~~~~ +\widetilde{p}_r^2 \left(\frac{39 \nu ^3}{16 \widetilde{r}^4}+\frac{9 \nu ^2}{2 \widetilde{r}^4}\right)+\widetilde{p}_r^4 \left(\frac{5 \nu ^2}{8 \widetilde{r}^3}-\frac{9 \nu ^3}{8 \widetilde{r}^3}\right)-\frac{101 \nu ^2}{56 \widetilde{r}^5}\nonumber\\
&~~~~~~~~~~~~~~~~~~~~ + \frac{1}{q} \Bigg(\widetilde{L}^4 \left(\frac{7 \nu ^3}{4 \widetilde{r}^7}-\frac{67 \nu ^2}{16 \widetilde{r}^7}+\frac{7 \nu }{8 \widetilde{r}^7}\right)+\widetilde{L}^2 \left(\widetilde{p}_r^2 \left(\frac{5 \nu ^3}{\widetilde{r}^5}-\frac{41 \nu ^2}{16 \widetilde{r}^5}+\frac{\nu }{4 \widetilde{r}^5}\right)-\frac{13 \nu ^3}{16 \widetilde{r}^6}-\frac{37 \nu ^2}{8 \widetilde{r}^6}+\frac{17 \nu }{2 \widetilde{r}^6}\right)\nonumber\\
&~~~~~~~~~~~~~~~~~~~~~~~~ +\widetilde{p}_r^2 \left(\frac{39 \nu ^3}{16 \widetilde{r}^4}+\frac{147 \nu ^2}{16 \widetilde{r}^4}-\frac{3 \nu }{\widetilde{r}^4}\right)+\widetilde{p}_r^4 \left(-\frac{23 \nu ^3}{16 \widetilde{r}^3}+\frac{13 \nu ^2}{8 \widetilde{r}^3}-\frac{5 \nu }{8 \widetilde{r}^3}\right)-\frac{5 \nu ^2}{2 \widetilde{r}^5}-\frac{33 \nu }{8 \widetilde{r}^5} \Bigg)
\Bigg\}\nonumber\\
&+\left((\widetilde{\textbf{S}}_{(1)}\cdot\widetilde{\textbf{r}})^2\right)\Bigg\{
\widetilde{L}^2 \left(\widetilde{p}_r^2 \left(-\frac{9 \nu ^3}{16 \widetilde{r}^7}-\frac{7 \nu ^2}{16 \widetilde{r}^7}\right)-\frac{9 \nu ^3}{8 \widetilde{r}^8}-\frac{25 \nu ^2}{8 \widetilde{r}^8}\right)-\frac{79 \nu ^3 \widetilde{p}_r^2}{16 \widetilde{r}^6}+\widetilde{p}_r^4 \left(\frac{9 \nu ^3}{8 \widetilde{r}^5}-\frac{5 \nu ^2}{8 \widetilde{r}^5}\right)+\frac{317 \nu ^2}{56 \widetilde{r}^7} \nonumber\\
&~~~~~~~~~~~~~~~~~~~~ +\frac{1}{q} \Bigg(\widetilde{L}^2 \left(\widetilde{p}_r^2 \left(-\frac{7 \nu ^3}{16 \widetilde{r}^7}-\frac{23 \nu ^2}{16 \widetilde{r}^7}+\frac{7 \nu }{16 \widetilde{r}^7}\right)-\frac{9 \nu ^3}{8 \widetilde{r}^8}-\frac{19 \nu ^2}{8 \widetilde{r}^8}+\frac{3 \nu }{8 \widetilde{r}^8}\right)\nonumber\\
&~~~~~~~~~~~~~~~~~~~~~~~~~ +\widetilde{p}_r^2 \left(-\frac{79 \nu ^3}{16 \widetilde{r}^6}-\frac{147 \nu ^2}{16 \widetilde{r}^6}+\frac{3 \nu }{\widetilde{r}^6}\right)+\widetilde{p}_r^4 \left(\frac{23 \nu ^3}{16 \widetilde{r}^5}-\frac{13 \nu ^2}{8 \widetilde{r}^5}+\frac{5 \nu }{8 \widetilde{r}^5}\right)+\frac{17 \nu ^2}{2 \widetilde{r}^7}+\frac{33 \nu }{8 \widetilde{r}^7}\Bigg)
\Bigg\}\nonumber\\
&+\left((\widetilde{\textbf{S}}_{(1)}\cdot\widetilde{\textbf{L}})^2\right)\Bigg\{
\widetilde{L}^2 \left(\frac{2 \nu ^2}{\widetilde{r}^7}-\frac{39 \nu ^3}{8 \widetilde{r}^7}\right)+\widetilde{p}_r^2 \left(\frac{29 \nu ^2}{16 \widetilde{r}^5}-\frac{123 \nu ^3}{8 \widetilde{r}^5}\right)+\frac{31 \nu ^3}{8 \widetilde{r}^6}+\frac{235 \nu ^2}{8 \widetilde{r}^6}\nonumber\\
&~~~~~~~~~~~~~~~~~~~~ +\frac{1}{q} \left(\widetilde{L}^2 \left(-\frac{35 \nu ^3}{8 \widetilde{r}^7}+\frac{175 \nu ^2}{16 \widetilde{r}^7}-\frac{2 \nu }{\widetilde{r}^7}\right)+\widetilde{p}_r^2 \left(-\frac{55 \nu ^3}{4 \widetilde{r}^5}+\frac{43 \nu ^2}{4 \widetilde{r}^5}-\frac{29 \nu }{16 \widetilde{r}^5}\right)+\frac{31 \nu ^3}{8 \widetilde{r}^6}+\frac{305 \nu ^2}{16 \widetilde{r}^6}-\frac{203 \nu }{8 \widetilde{r}^6}\right)
\Bigg\}\nonumber\\
&+\left((\widetilde{\textbf{S}}_{(1)}\cdot\widetilde{\textbf{r}})(\widetilde{\textbf{S}}_{(1)}\cdot\widetilde{\textbf{L}})\right) \widetilde{p}_r\Bigg\{
\widetilde{L}^2 \left(\frac{7 \nu ^2}{16 \widetilde{r}^7}-\frac{39 \nu ^3}{16 \widetilde{r}^7}\right)+ \left(\frac{13 \nu ^3}{4 \widetilde{r}^6}+\frac{13 \nu ^2}{4 \widetilde{r}^6}\right)+\widetilde{p}_r^2 \left(\frac{\nu ^2}{16 \widetilde{r}^5}-\frac{15 \nu ^3}{4 \widetilde{r}^5}\right)\nonumber\\
&~~~~~~~~~~~~~~~~~~~~~~~~~~~~~~~~~ +\frac{1}{q} \bigg(\widetilde{L}^2 \left(-\frac{35 \nu ^3}{16 \widetilde{r}^7}+\frac{11 \nu ^2}{4 \widetilde{r}^7}-\frac{7 \nu }{16 \widetilde{r}^7}\right)+ \left(\frac{13 \nu ^3}{4 \widetilde{r}^6}+\frac{7 \nu ^2}{\widetilde{r}^6}-\frac{25 \nu }{4 \widetilde{r}^6}\right)\nonumber\\
&~~~~~~~~~~~~~~~~~~~~~~~~~~~~~~~~~~~~~~~ +\widetilde{p}_r^2 \left(-\frac{25 \nu ^3}{8 \widetilde{r}^5}+\frac{19 \nu ^2}{8 \widetilde{r}^5}-\frac{\nu }{16 \widetilde{r}^5}\right)\bigg)
\Bigg\}\nonumber\\
&+(1\leftrightarrow 2)
\end{align}

\subsection{$\rm ES^2$ sector up to \NNLO}

\begin{align}
\widetilde{\mathcal{H}}^{\rm ES^2}_{\text{LO}}= 
\left(C^{(0)}_{\rm ES^2}\right)_{(1)}\left(\widetilde{\textbf{S}}_{(1)}\cdot\widetilde{\textbf{S}}_{(1)}\right)\left\{
-\frac{1}{q}\frac{\nu }{2 \widetilde{r}^3}
\right\}
+\left(C^{(0)}_{\rm ES^2}\right)_{(1)}\left((\widetilde{\textbf{S}}_{(1)}\cdot\widetilde{r})^2\right)\left\{
\frac{1}{q}\frac{3 \nu}{2 \widetilde{r}^5}
\right\} +(1\leftrightarrow 2)
\end{align}

\begin{align}
\widetilde{\mathcal{H}}^{\rm ES^2}_{\text{NLO}}&= 
\left(C^{(0)}_{\rm ES^2}\right)_{(1)}\left(\widetilde{\textbf{S}}_{(1)}\cdot\widetilde{\textbf{S}}_{(1)}\right)\left\{
-\frac{\widetilde{L}^2 \nu ^2}{2 \widetilde{r}^5}+\frac{\nu ^2 \widetilde{p}_r^2}{\widetilde{r}^3}-\frac{\nu ^2}{2 \widetilde{r}^4}+ \frac{1}{q} \left(\widetilde{L}^2 \left(-\frac{3 \nu ^2}{4 \widetilde{r}^5}-\frac{5 \nu }{4 \widetilde{r}^5}\right)+\frac{\nu  \widetilde{p}_r^2}{4 \widetilde{r}^3}-\frac{\nu ^2}{2 \widetilde{r}^4}+\frac{2 \nu }{\widetilde{r}^4}\right)
\right\}\nonumber\\
&+\left(C^{(0)}_{\rm ES^2}\right)_{(1)}\left((\widetilde{\textbf{S}}_{(1)}\cdot\widetilde{\textbf{r}})^2\right)\left\{
\frac{5 \nu ^2}{2 \widetilde{r}^6}-\frac{\nu ^2 \widetilde{p}_r^2}{\widetilde{r}^5}+\frac{1}{q} \left(\widetilde{L}^2 \left(\frac{3 \nu ^2}{4 \widetilde{r}^7}+\frac{9 \nu }{4 \widetilde{r}^7}\right)+\widetilde{p}_r^2 \left(\frac{2 \nu ^2}{\widetilde{r}^5}-\frac{3 \nu }{4 \widetilde{r}^5}\right)+\frac{5 \nu ^2}{2 \widetilde{r}^6}-\frac{6 \nu }{\widetilde{r}^6}\right)
\right\}\nonumber\\
&+\left(C^{(0)}_{\rm ES^2}\right)_{(1)}\left((\widetilde{\textbf{S}}_{(1)}\cdot\widetilde{\textbf{L}})^2\right)\left\{
\frac{\nu ^2}{2 \widetilde{r}^5}+ \frac{1}{q} \left(\frac{\nu ^2}{2 \widetilde{r}^5}+\frac{3 \nu }{2 \widetilde{r}^5}\right)
\right\}\nonumber\\
&+\left(C^{(0)}_{\rm ES^2}\right)_{(1)}\left((\widetilde{\textbf{S}}_{(1)}\cdot\widetilde{\textbf{r}})(\widetilde{\textbf{S}}_{(1)}\cdot\widetilde{\textbf{L}})\right) \widetilde{p}_r \left\{
-\frac{\nu ^2}{2 \widetilde{r}^5}+ \frac{1}{q} \left(-\frac{2 \nu ^2}{\widetilde{r}^5}-\frac{3 \nu }{2 \widetilde{r}^5}\right)
\right\}\nonumber\\
&+(1\leftrightarrow 2)
\end{align}

\begin{align}
\widetilde{\mathcal{H}}^{\rm ES^2}_{\rm N^2LO}&= 
\left(C^{(0)}_{\rm ES^2}\right)_{(1)}\left(\widetilde{\textbf{S}}_{(1)}\cdot\widetilde{\textbf{S}}_{(1)}\right)\Bigg\{
\widetilde{L}^4 \left(\frac{\nu ^2}{4 \widetilde{r}^7}-\frac{3 \nu ^3}{4 \widetilde{r}^7}\right)+\widetilde{L}^2 \left(\widetilde{p}_r^2 \left(-\frac{3 \nu ^3}{2 \widetilde{r}^5}-\frac{\nu ^2}{4 \widetilde{r}^5}\right)+\frac{\nu ^3}{4 \widetilde{r}^6}+\frac{\nu ^2}{2 \widetilde{r}^6}\right) \nonumber\\
&~~~~~~~~~~~~~~~~~~~~ + \widetilde{p}_r^2 \left(-\frac{7 \nu ^3}{4 \widetilde{r}^4}-\frac{15 \nu ^2}{4 \widetilde{r}^4}\right)+\widetilde{p}_r^4 \left(\frac{3 \nu ^3}{\widetilde{r}^3}-\frac{\nu ^2}{2 \widetilde{r}^3}\right) + \frac{27 \nu ^2}{14 \widetilde{r}^5}\nonumber\\
&~~~~~~~~~~~~~~~~~~~~ + \frac{1}{q} \Bigg(\widetilde{L}^4 \left(-\frac{15 \nu ^3}{16 \widetilde{r}^7}-\frac{7 \nu ^2}{4 \widetilde{r}^7}+\frac{9 \nu }{16 \widetilde{r}^7}\right)+\widetilde{L}^2 \left(\widetilde{p}_r^2 \left(-\frac{9 \nu ^3}{4 \widetilde{r}^5}-\frac{17 \nu ^2}{4 \widetilde{r}^5}+\frac{3 \nu }{8 \widetilde{r}^5}\right)+\frac{\nu ^3}{4 \widetilde{r}^6}+\frac{49 \nu ^2}{8 \widetilde{r}^6}+\frac{15 \nu }{2 \widetilde{r}^6}\right)\nonumber\\
&~~~~~~~~~~~~~~~~~~~~~~~~ +\widetilde{p}_r^2 \left(-\frac{7 \nu ^3}{4 \widetilde{r}^4}+\frac{27 \nu ^2}{8 \widetilde{r}^4}-\frac{3 \nu }{2 \widetilde{r}^4}\right)+\widetilde{p}_r^4 \left(\frac{3 \nu ^3}{2 \widetilde{r}^3}+\frac{5 \nu ^2}{4 \widetilde{r}^3}-\frac{3 \nu }{16 \widetilde{r}^3}\right)+\frac{\nu ^2}{4 \widetilde{r}^5}-\frac{19 \nu }{4 \widetilde{r}^5} \Bigg)
\Bigg\}\nonumber\\
&+\left(C^{(0)}_{\rm ES^2}\right)_{(1)}\left((\widetilde{\textbf{S}}_{(1)}\cdot\widetilde{\textbf{r}})^2\right)\Bigg\{
\widetilde{L}^2 \left(\widetilde{p}_r^2 \left(-\frac{3 \nu ^3}{4 \widetilde{r}^7}-\frac{\nu ^2}{8 \widetilde{r}^7}\right)+\frac{5 \nu ^3}{4 \widetilde{r}^8}+\frac{17 \nu ^2}{2 \widetilde{r}^8}\right)+\widetilde{p}_r^2 \left(\frac{27 \nu ^3}{4 \widetilde{r}^6}+\frac{15 \nu ^2}{4 \widetilde{r}^6}\right)\nonumber\\
&~~~~~~~~~~~~~~~~~~~~ +\widetilde{p}_r^4 \left(\frac{5 \nu ^2}{2 \widetilde{r}^5}-\frac{3 \nu ^3}{\widetilde{r}^5}\right)-\frac{123 \nu ^2}{14 \widetilde{r}^7} \nonumber\\
&~~~~~~~~~~~~~~~~~~~~ +\frac{1}{q} \Bigg(\widetilde{L}^4 \left(\frac{9 \nu ^3}{16 \widetilde{r}^9}+\frac{3 \nu ^2}{\widetilde{r}^9}-\frac{15 \nu }{16 \widetilde{r}^9}\right)+\widetilde{L}^2 \left(\widetilde{p}_r^2 \left(\frac{9 \nu ^3}{8 \widetilde{r}^7}+\frac{23 \nu ^2}{4 \widetilde{r}^7}-\frac{3 \nu }{4 \widetilde{r}^7}\right)+\frac{5 \nu ^3}{4 \widetilde{r}^8}-\frac{\nu ^2}{4 \widetilde{r}^8}-\frac{27 \nu }{2 \widetilde{r}^8}\right)\nonumber\\
&~~~~~~~~~~~~~~~~~~~~~~~~~ +\widetilde{p}_r^2 \left(\frac{27 \nu ^3}{4 \widetilde{r}^6}-\frac{177 \nu ^2}{8 \widetilde{r}^6}+\frac{9 \nu }{2 \widetilde{r}^6}\right)+\widetilde{p}_r^4 \left(\frac{3 \nu ^3}{2 \widetilde{r}^5}-\frac{7 \nu ^2}{4 \widetilde{r}^5}+\frac{9 \nu }{16 \widetilde{r}^5}\right)+\frac{57 \nu }{4 \widetilde{r}^7}-\frac{23 \nu ^2}{4 \widetilde{r}^7}\Bigg)
\Bigg\}\nonumber\\
&+\left(C^{(0)}_{\rm ES^2}\right)_{(1)}\left((\widetilde{\textbf{S}}_{(1)}\cdot\widetilde{\textbf{L}})^2\right)\Bigg\{
\widetilde{L}^2 \left(\frac{3 \nu ^3}{4 \widetilde{r}^7}-\frac{5 \nu ^2}{4 \widetilde{r}^7}\right)+\widetilde{p}_r^2 \left(\frac{9 \nu ^3}{4 \widetilde{r}^5}+\frac{11 \nu ^2}{8 \widetilde{r}^5}\right)-\frac{\nu ^3}{2 \widetilde{r}^6}-\frac{5 \nu ^2}{2 \widetilde{r}^6}\nonumber\\
&~~~~~~~~~~~~~~~~~~~~ +\frac{1}{q} \left(\widetilde{L}^2 \left(\frac{3 \nu ^3}{4 \widetilde{r}^7}+\frac{5 \nu ^2}{4 \widetilde{r}^7}-\frac{3 \nu }{4 \widetilde{r}^7}\right)+\widetilde{p}_r^2 \left(\frac{21 \nu ^3}{8 \widetilde{r}^5}+\frac{8 \nu ^2}{\widetilde{r}^5}-\frac{3 \nu }{8 \widetilde{r}^5}\right)-\frac{\nu ^3}{2 \widetilde{r}^6}-\frac{65 \nu ^2}{8 \widetilde{r}^6}-\frac{9 \nu }{\widetilde{r}^6}\right)
\Bigg\}\nonumber\\
&+\left(C^{(0)}_{\rm ES^2}\right)_{(1)}\left((\widetilde{\textbf{S}}_{(1)}\cdot\widetilde{\textbf{r}})(\widetilde{\textbf{S}}_{(1)}\cdot\widetilde{\textbf{L}})\right) \widetilde{p}_r\Bigg\{
-\frac{11 \widetilde{L}^2 \nu ^2}{8 \widetilde{r}^7}+ \left(\frac{25 \nu ^2}{\widetilde{r}^6}-\frac{7 \nu ^3}{4 \widetilde{r}^6}\right)+\widetilde{p}_r^2 \left(\frac{31 \nu ^2}{8 \widetilde{r}^5}-\frac{3 \nu ^3}{4 \widetilde{r}^5}\right)\nonumber\\
&~~~~~~~~~~~~~~~~~~~~~~~~~~~~~~~~~ +\frac{1}{q} \bigg(\widetilde{L}^2 \left(-\frac{9 \nu ^3}{8 \widetilde{r}^7}-\frac{11 \nu ^2}{4 \widetilde{r}^7}+\frac{3 \nu }{8 \widetilde{r}^7}\right)+ \left(-\frac{7 \nu ^3}{4 \widetilde{r}^6}+\frac{42 \nu ^2}{\widetilde{r}^6}+\frac{9 \nu }{\widetilde{r}^6}\right)\nonumber\\
&~~~~~~~~~~~~~~~~~~~~~~~~~~~~~~~~~~~~~~~ +\widetilde{p}_r^2 \left(-\frac{39 \nu ^3}{8 \widetilde{r}^5}-\frac{\nu ^2}{2 \widetilde{r}^5}+\frac{9 \nu }{8 \widetilde{r}^5}\right)\bigg)
\Bigg\}\nonumber\\
&+(1\leftrightarrow 2)
\end{align}

\bibliographystyle{JHEP}
\bibliography{biblio}

\end{document}